\documentclass[11pt,a4paper]{article}
\usepackage{jheppub}
\usepackage{graphics,graphicx,lineno}
\usepackage{epstopdf}
\usepackage{draftcopy}
\usepackage{color}
\usepackage{multirow}
\title{New results on \boldmath{$\nu_\mu \to\nu_\tau$} appearance with the OPERA experiment in the CNGS beam}

\collaborationImg{\includegraphics[width=2.5cm]{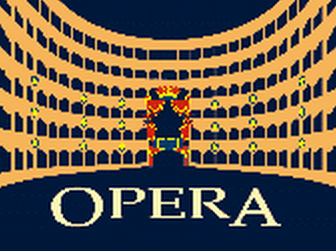}~~~The OPERA Collaboration}

\author[a]{N.~Agafonova,}
\author[b]{A.~Aleksandrov,}
\author[c]{A.~Anokhina,}
\author[d]{S.~Aoki,}
\author[e]{A.~Ariga,}
\author[e]{T.~Ariga,}
\author[x]{T.~Asada,}
\author[f]{D.~Autiero,}
\author[g]{A.~Badertscher,}
\author[e]{A.~Ben Dhahbi,}
% ...

\affiliation[a]{INR Institute for Nuclear Research, Russian Academy of Sciences RUS-117312, Moscow, Russia}
\affiliation[b]{INFN Sezione di Napoli, I-80125 Napoli, Italy}
\affiliation[c]{SINP MSU-Skobeltsyn Institute of Nuclear Physics, Lomonosov Moscow State University, \mbox{RUS-119992} Moscow, Russia}
\affiliation[d]{Kobe University, J-657-8501 Kobe, Japan}
\affiliation[e]{Albert Einstein Center for Fundamental Physics, Laboratory for High Energy Physics (LHEP), University of Bern, CH-3012 Bern, Switzerland}
\affiliation[f]{IPNL, Universit\'e Claude Bernard Lyon 1, CNRS/IN2P3, F-69622 Villeurbanne, France}
\affiliation[g]{ETH Zurich, Institute for Particle Physics, CH-8093 Zurich, Switzerland}
% ...
\affiliation[x]{Nagoya University, J-464-8602 Nagoya, Japan}
% ...

\emailAdd{umut.kose@cern.ch} \emailAdd{andrea.longhin@lnf.infn.it}
\abstract{
The \textsc{OPERA} neutrino experiment is designed to perform the
first observation of neutrino oscillations in direct appearance mode
in the \mbox{$\nu_\mu \to \nu_\tau$} channel, via the detection of the
$\tau$-leptons created in charged current \mbox{$\nu_\tau$
  interactions}.  The detector, located in the underground Gran Sasso
Laboratory, consists of an emulsion/lead target with an average mass
of about 1.2~kt, complemented by electronic detectors. It is exposed
to the \CERN~Neutrinos to Gran Sasso beam, with a baseline of 730 km
and a mean energy of 17~GeV.  The observation of the first $\nu_\tau$
candidate event and the analysis of the 2008-2009 neutrino sample have
been reported in previous publications.  This work describes
substantial improvements in the analysis and in the evaluation of the
detection efficiencies and backgrounds using new simulation tools.
The analysis is extended to a sub-sample of 2010 and 2011 data,
resulting from an electronic detector-based pre-selection, in which an
additional $\nu_\tau$ candidate has been observed.  The significance
of the two events in terms of a $\nu_\mu\to\nu_\tau$ oscillation
signal is of \SigmaB.
} 


\arxivnumber{xxxx.xxxx}

\newcommand{\OPERA}{\textsc{OPERA}}
\newcommand{\SK}{\textsc{Super-Kamiokande}}
\newcommand{\DOUBLECHOOZ}{\textsc{Double Chooz}}
\newcommand{\DAYABAY}{\textsc{Daya Bay}}
\newcommand{\RENO}{\textsc{RENO}}
\newcommand{\TTWOK}{\textsc{T2K}}
\newcommand{\CERN}{\textsc{CERN}}
\newcommand{\CNGS}{\textsc{CNGS}}
\newcommand{\CHORUS}{\textsc{CHORUS}}
\newcommand{\WANF}{\textsc{WANF}}
\newcommand{\FLUKA}{\textsc{FLUKA}}
\newcommand{\GENIE}{\textsc{GENIE}}
\newcommand{\FinalPOT}{{\color{black}{$17.97 \times 10^{19}$}} }
\newcommand{\COnehfinal}{{\color{black}{$0.027 \pm 0.005$}} }
\newcommand{\CThreehfinal}{{\color{black}{$0.12 \pm 0.02$}} }
\newcommand{\CElecfinal}{{\color{black}{$0.020 \pm 0.004$}} }
\newcommand{\CMuonfinal}{{\color{black}{$0.012 \pm 0.005$}} }
\newcommand{\charmuncertainty}{{\color{black}{20\%}}}
\newcommand{\cOnehsuppr}{{\color{black}{$3.3\times 10^{-5}$}} }
\newcommand{\cThreehsuppr}{{\color{black}{$3.4\times10^{-4}$}} }
\newcommand{\HOneThreefinal}{{\color{black}{$0.018 \pm 0.005$}} }
\newcommand{\LASfinal}{{\color{black}{$0.009 \pm 0.005$}} }
\newcommand{\ALLfinal}{{\color{black}{$3.2 \pm 0.3$}} }
\newcommand{\ALLTHISfinal}{{\color{black}{$1.53 \pm 0.16$}} }
\newcommand{\ALLBTHISfinal}{{\color{black}{$0.18 \pm 0.02$}} }
\newcommand{\SigmaA}{{\color{black}{$2.20\sigma$}} }
\newcommand{\SigmaB}{{\color{black}{$2.40\sigma$}}}
\newcommand{\SigmaOneh}{{\color{black}{$1.94\sigma$}} }
\newcommand{\SigmaThreeh}{{\color{black}{$1.23\sigma$}} }
\newcommand{\pValueOneh}{{\color{black}{$p_{1h} = 2.6 \times 10^{-2}$}} }
\newcommand{\pValueThreeh}{{\color{black}{$p_{3h} = 1.10 \times 10^{-1}$}} }
\newcommand{\pValueSum}{{\color{black}{$p=1.36\times 10^{-2}$}} }
\newcommand{\NED}{{\color{black}{11149}}}
\newcommand{\NPRED}{{\color{black}{9267}}}
\newcommand{\NBRICK}{{\color{black}{4135}}}
\newcommand{\NDS}{{\color{black}{3969}}}
\newcommand{\NOTDSONLOC}{{\color{black}{4}}\%}
\newcommand{\FRACTEN}{{\color{black}{78}}\%} 
\newcommand{\FRACELEVEN}{{\color{black}{33}}\%} 

\begin{document}
\maketitle
\flushbottom


\section{Introduction}

Flavour transitions between neutrino species were predicted nearly 50
years ago \cite{ref1_a,ref1_b,ref1_c}. In 1998 the \SK~experiment
observed a strong deficit of atmospheric muon neutrinos in the data,
and interpreted it as a result of such transitions
\cite{ref2_a,ref2_b,ref2_c}. A few years later, after being
investigated for decades by real-time and geochemical
experiments~\cite{sol_a,sol_b,sol_c,sol_d,sol_e,sol_f}, the so-called
solar neutrino problem could also be interpreted (after the
\textsc{SNO} results~\cite{SNO}) in terms of neutrino oscillations in
presence of the \textsc{MSW} effect~\cite{MSW} in the Sun's
matter. The disappearance of $\nu_\mu$ in atmospheric neutrinos
\cite{ref2_a,ref2_b,ref2_c,atm_a,atm_b,atm_c} was confirmed at
\mbox{accelerator-based} long baseline
experiments~\cite{ref5_a,ref5_b,ref5_c}. The observation of an
appearance of neutrinos from oscillations consistent with the
disappearance results is still missing. The \OPERA~experiment
\cite{ref8,proposal} has the capability of detecting the appearance of
a small $\nu_\tau$ component in a $\nu_\mu$ beam.  The fact that the
\mbox{$\nu_\mu \to\nu_e$} oscillation could not be the dominant reason
of the deficit of $\nu_\mu$ was also established by the measurement of
atmospheric $\nu_e$ rates and confirmed by nuclear reactor experiments
at short baselines~\cite{ref6_a,ref6_b} in the late nineties. Only
recently, the probability amplitude of the \mbox{$\nu_\mu\to \nu_e$}
transition, governed by the $\theta_{13}$ mixing angle, has been
measured at a long baseline neutrino beam (\TTWOK~\cite{ref7}). The
first evidence of the disappearance of $\bar{\nu}_e$ has come from
reactor experiments at short baselines and $\theta_{13}$ has been
measured (\DAYABAY~\cite{citDB}, \DOUBLECHOOZ~\cite{citDC},
\RENO~\cite{citRE}).  \SK~also recently reported a statistical
evidence of $\nu_\mu\to \nu_\tau$ transitions \cite{SKapp} in its
atmospheric neutrino sample but with a modest signal to background
noise due to the difficulty in cleanly reconstructing $\nu_\tau$
events in a water-Cherenkov detector.  A positive evidence from
\OPERA~would definitely prove that the $\nu_\mu\to \nu_\tau$
transition is the reason of $\nu_\mu$ deficit and it is the dominant
mechanism at the atmospheric scale, providing an essential constraint
for the establishment of the 3-flavor mixing scheme. Furthermore the
measurement can constrain phenomenological models such as neutrino Non
Standard Interactions~(\textsc{NSI})~\cite{NSIopera_a,NSIopera_b} or
sterile neutrinos, which, for certain choices of the parameters,
result in modifications to the expected $\nu_\tau$ appearance rate.

After two years of technical and low-intensity runs (2006-2007) the
\OPERA~detector in the Gran Sasso underground laboratory
(\textsc{LNGS}) has been exposed from 2008 to 2012 to the high-energy
\CERN~Neutrinos to Gran Sasso (\CNGS) beam~\cite{ref10}. The detection
of the first \CNGS~neutrino interactions in \OPERA~was reported
in~\cite{ref11_a,ref11_b} and the observation of a first $\nu_\tau$
candidate event was presented in~\cite{ref1sttau}, using a sample
corresponding to $1.9 \times 10^{19}$ protons on target (pot). An
update based on the data collected in years 2008-09
($4.88\times10^{19}$~pot) was presented in~\cite{ref2ndoscpaper}.
This paper presents the progress of the $\nu_\tau$ appearance search,
through a re-evaluation of the efficiencies and of the backgrounds
based on a full simulation of the complete analysis chain. Moreover a
larger data sample has been exploited which allowed the observation of
a second $\nu_\tau$ candidate event.

After an introduction explaining the operation of the experiment
(Sect.~\ref{sec:det}), the data selection (Sect.~\ref{sec:evsel}) and
the simulation (Sect.~\ref{sec:sim}), the results
(Sect.~\ref{sec:res}) will be described, giving particular emphasis to
the second $\nu_{\tau}$ candidate.  The significance of the
observation of two candidates given the present statistics and the
estimated level of background is addressed in Sect.~\ref{sec:multiv}
in terms of exclusion of the null-hypothesis.

\section{Detector and neutrino beam}
\label{sec:det}
\begin{figure}
\centering
\includegraphics[width=10cm]{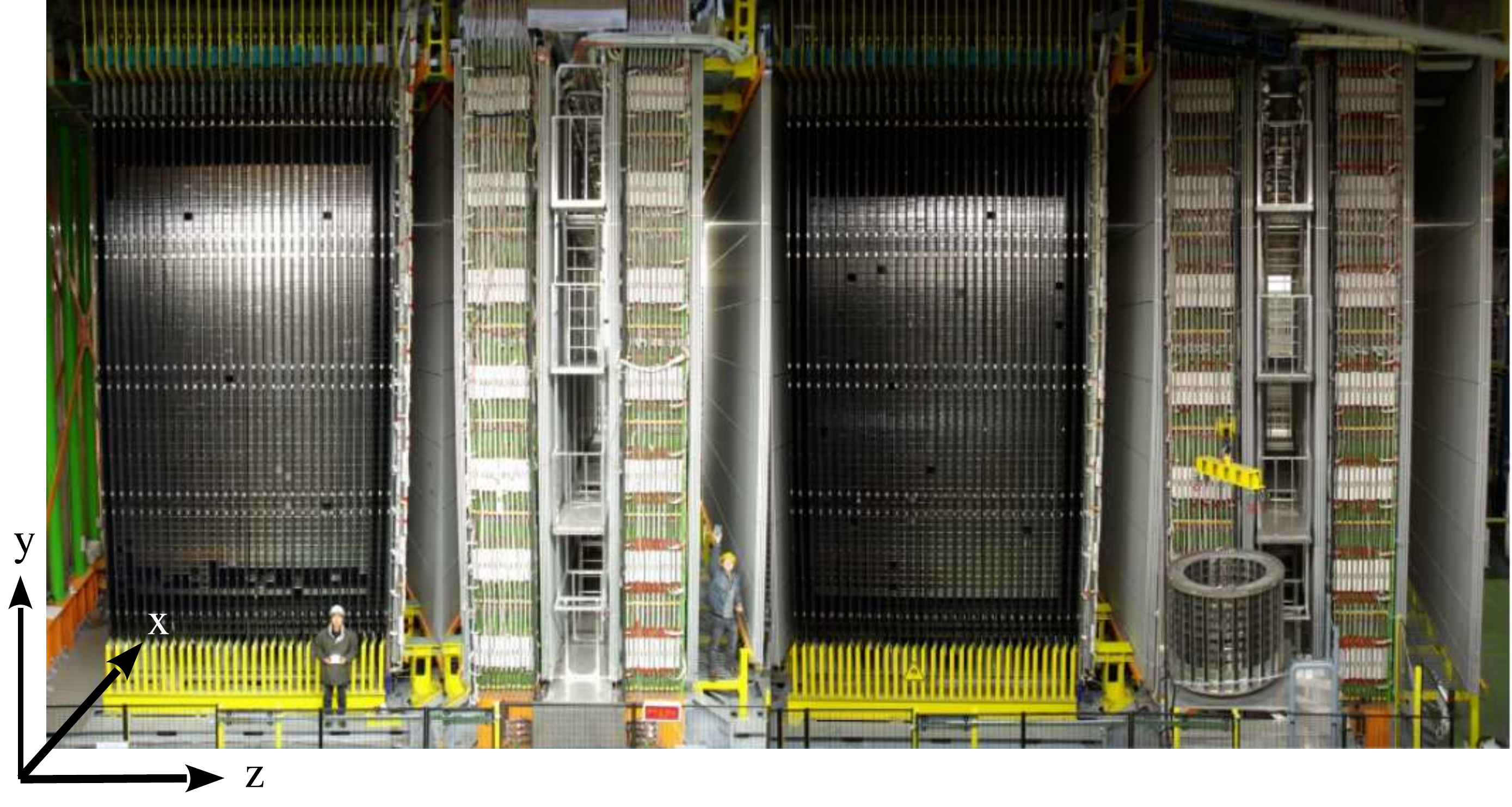}
\caption{A picture of the \OPERA~detector. \CNGS~neutrinos travel from
  left to right.  The (right-handed) reference frame is oriented such
  that: the $y$-axis is perpendicular to the hall floor and pointing
  up; the $z$-axis is orthogonal to the brick walls and oriented as
  the incoming neutrinos.  The angle between the neutrino direction
  and the $z$-axis projected into the $yz$ plane is of 58~mrad.}
\label{fig:operadet}
\end{figure}

The \OPERA~detector \cite{ref13} is designed to tackle a challenging
task: achieving micro-metric tracking accuracy over a very large
detector volume spanning about \mbox{(6.5 $\times$ 6.5 $\times$
  8)~m$^3$}. The scale of the required granularity is set by the
flight length of $\tau$ leptons, which for the \CNGS~beam has a
roughly exponential distribution with a mean of about 600~$\mu$m.
This challenge was addressed by using nuclear emulsion based trackers.
Another important constraint is related to the practical impossibility
to analyze the full emulsion surface ${\mathcal{O}}$(0.1~km$^2$), even
with the state-of-the art automatic scanning technology. This,
together with other constraints, resulted in a highly modular target
made of units based on the Emulsion Cloud Chamber (ECC) technique,
hereafter called bricks, interspersed with pair of planes of
horizontal and vertical scintillator strips (called Target Tracker or
TT~\cite{ref15_c}) that allow locating with a centimetric resolution
the unit in which the neutrino interaction occurred.  A brick is
composed of 57 emulsion films interleaved with 56, 1 mm thick, lead
plates for a mass of 8.3 kg. Its thickness along the beam direction
corresponds to about 10 radiation lengths and its transverse size is
\mbox{128 $\times$ 102 mm$^2$}. A film consists of two 44~$\mu$m
layers deposited on each side of a 205~$\mu$m plastic base.  Another
key ingredient for the experiment are the Changeable Sheet (CS)
doublets \cite{CSpaper}, attached to the downstream face of each
brick.  This is a pair of films having received in the underground
laboratory a special treatment (refreshing) aiming at erasing most of
the cosmic background accumulated since their fabrication.  Their
scanning allows a relatively fast feedback on the prediction of the
electronic detectors (ED) and provides a prediction of the event
position in the brick at the ${\mathcal{O}}(10) \mu$m level, thus
greatly helping the vertex location.  Finally a magnetic spectrometer
system instrumented with Resistive Plate Chamber (RPC) detectors and
high-precision Drift Tubes (DT), is used for the task of identifying
muons and measuring their charge and momentum.  A good muon
identification capability is essential to reduce the background to
$\tau$ decays from charmed particles produced in charged current (CC)
$\nu_\mu$ interactions.  The detector (Fig.~\ref{fig:operadet}) is
divided into two identical units called Super Modules (SM), each
consisting of a target and a spectrometer section. The average number
of bricks has been about 140000 for a target mass of about 1.2~kt.

\OPERA~was exposed to the \CNGS~$\nu_\mu$ beam~\cite{ref10} at a
long-baseline, 730 km away from the source.  The neutrino beam,
produced by 400 GeV-protons accelerated in the SPS, has an average
energy of about 17 GeV, optimised for the observation of $\nu_\tau$ CC
interactions in the \OPERA~detector.  In terms of interactions, the
$\bar{\nu}_\mu$ contamination is 2.1\%, the $\nu_e$ and $\bar{\nu}_e$
contaminations are together below 1\%, while the intrinsic $\nu_\tau$
component (from $D_s$ decays in the \CNGS~target and beam-dump) is of
${\mathcal{O}}(10^{-6})$, hence negligible.

\section{Data samples and event selection}
\label{sec:evsel}
\CNGS~completed its operation on \mbox{December 3, 2012}.  A sample
corresponding to \FinalPOT pot has been registered by the detector
since the beginning of the program in 2008.  The first task of the
electronic detectors DAQ \cite{operadaq} is the time-tagging of the
hits allowing the selection of events in coincidence with the two
10.5~$\mu$s-wide \CNGS~spills separated by 50~ms (the so-called
``on-time'' events).

An on-line filtering (TT local trigger) is then applied in order to
remove background from random noise in the detector: hits are required
in the horizontal and vertical views of at least two planes or the
presence of at least 4 hits in a single plane is required with the sum
of their photomultiplier ADC signals exceeding 500 counts
(corresponding to about ten photo-electrons)\footnote{The cut
  parameters were 10 hits and 30 photo-electrons for 2008 and 2009
  runs.}.

The event is then classified as being CC-like (hereafter ``1$\mu$'')
or neutral current (NC) like (``0$\mu$'') using the data of the target
tracker and the spectrometers. Recorded hits are processed by a
pattern recognition algorithm and sub-samples of hits in both views
are grouped into three dimensional (3D) tracks.  A 3D-track is tagged
as a muon if the product of its length and the density along its path
is larger than 660~g/cm$^{2}$.  An event is classified as 1$\mu$ if
either it contains at least one 3D-track tagged as a
muon~\cite{OPERAEDpaper} or the total number of TT and RPC planes
having at least one hit is larger than 19.  The complementary sample
is defined as~0$\mu$.  About 19\% of the NC events are classified as
$1\mu$ while only 6\% of CC events are classified as $0\mu$.  The
momentum of 3D-tracks is calculated from their bending in the
spectrometer magnetic field and/or their range with a Kalman
filter-based reconstruction algorithm.

A pure sample of 106422 \CNGS~on-time neutrino interactions is
selected (Tab.~\ref{tab:samples}, 3$^{\rm{rd}}$ column).  About 60\%
of this sample results from neutrino interactions in the rock in front
of the detector typically producing long passing-through muon tracks,
while the rest is from interactions occurring, in about equal
fractions, in the target (contained events) and in the spectrometer.
\begin{table}
\centering
\small
\begin{tabular}{ccccc}
\hline
year & pot (10$^{19}$)& on-time events & contained events& average No. of bricks\\
\hline
2008 & 1.74 & 10141 & 1931  & 141475\\
2009 & 3.53 & 21455 & 4005  & 147344\\
2010 & 4.09 & 25497 & 4515  & 144398\\
2011 & 4.75 & 28195 & 5131  & 138798\\
2012 & 3.86 & 21134 & 3923  & 135142\\
\hline
all & 17.97 &  106422 & 19505 & 141431\\
\hline 
\end{tabular}
\caption{Summary of the collected data samples. The given pot are corrected
  for the live-time of the detector. The meaning of the other columns
  is detailed in the text.}
\label{tab:samples}
\end{table}
A classifier algorithm, \textsc{OpCarac}~\cite{refopcarac}, is applied
to select the contained events yielding a total of
{\color{black}{19505}} interactions (Tab.~\ref{tab:samples},
4$^{\rm{th}}$~column).  This number corresponds to an average rate of
about 18 contained neutrino interactions per day which have been
stably recorded with an overall \mbox{dead-time} of the data
acquisition and detectors of 1.5\%.

Contained events are processed by a brick-finding algorithm
\cite{BFref}.  The topology and the energy deposition in the TT
scintillator strips, as well as the muon track information (when
available) are used to define a three-dimensional probability density
map for the vertex position. This probability is integrated over the
volumes of the bricks and these are ranked in order of decreasing
probability, for extraction and analysis.  In the following, the
highest-probability brick will be denoted as HPB.

\subsection{Brick pre-selection and choice of the analysed sample}
\label{sec:presel}
The analysis of the 2008-2009 inclusive sample (30\% of the overall
number of pot) was reported in~\cite{ref2ndoscpaper}.  In that sample
of 2738 fully analysed events one event was recognised as a $\nu_\tau$
candidate decaying to a single charged hadron (noted as $\tau\to
1h$)\cite{ref1sttau}. More precisely the measurement of the final
state topology and kinematics strongly favors this interpretation:
$\tau^-\to\nu_\tau\rho^-$ with $\rho^-\to \pi^-\pi^0$ and
$\pi^0\to\gamma\gamma$.

Since then the scanning strategy has been modified to accelerate the
finding of a significant signal for $\nu_\mu\to \nu_\tau$ oscillation.
Priority was given to the scanning of a pre-selected sample of bricks:

\begin{itemize}
\item 2008-2009 data: analysis of up to two bricks per event without event pre-selection.
\item 2010 data: analysis of the HPBs, for all $0\mu$ events and for $1\mu$ events
  with a muon momentum $p_\mu < 15$~GeV/$c$.
\item 2011 data: analysis of the HPBs for all $0\mu$ events.
\end{itemize}
This pre-selection is temporary and the analysis of lower-priority
bricks will occur in the near future.

$0\mu$ events form a preferred sample to search for $\tau$ decays
since these essentially contain signals of the electron final state,
with a branching ratio of \mbox{$(17.85 \pm 0.05)$\%}, and of the
1-prong, \mbox{$(49.52 \pm 0.07)$}\% and 3-prong, \mbox{$(15.19 \pm
  0.08)$}\% hadronic channels.

\begin{figure}
\centering
\includegraphics[width=7.5cm]{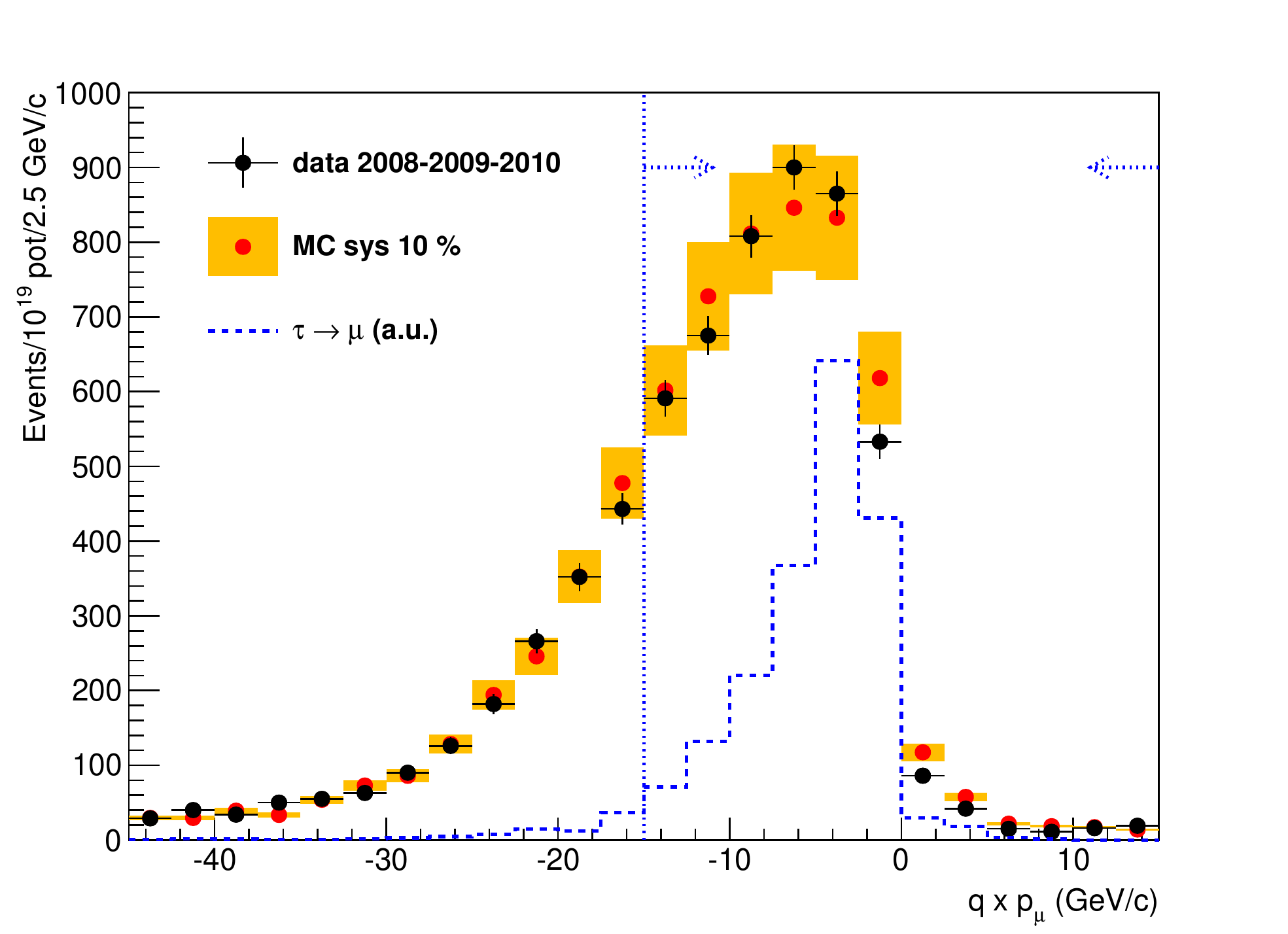}%
\includegraphics[width=7.5cm]{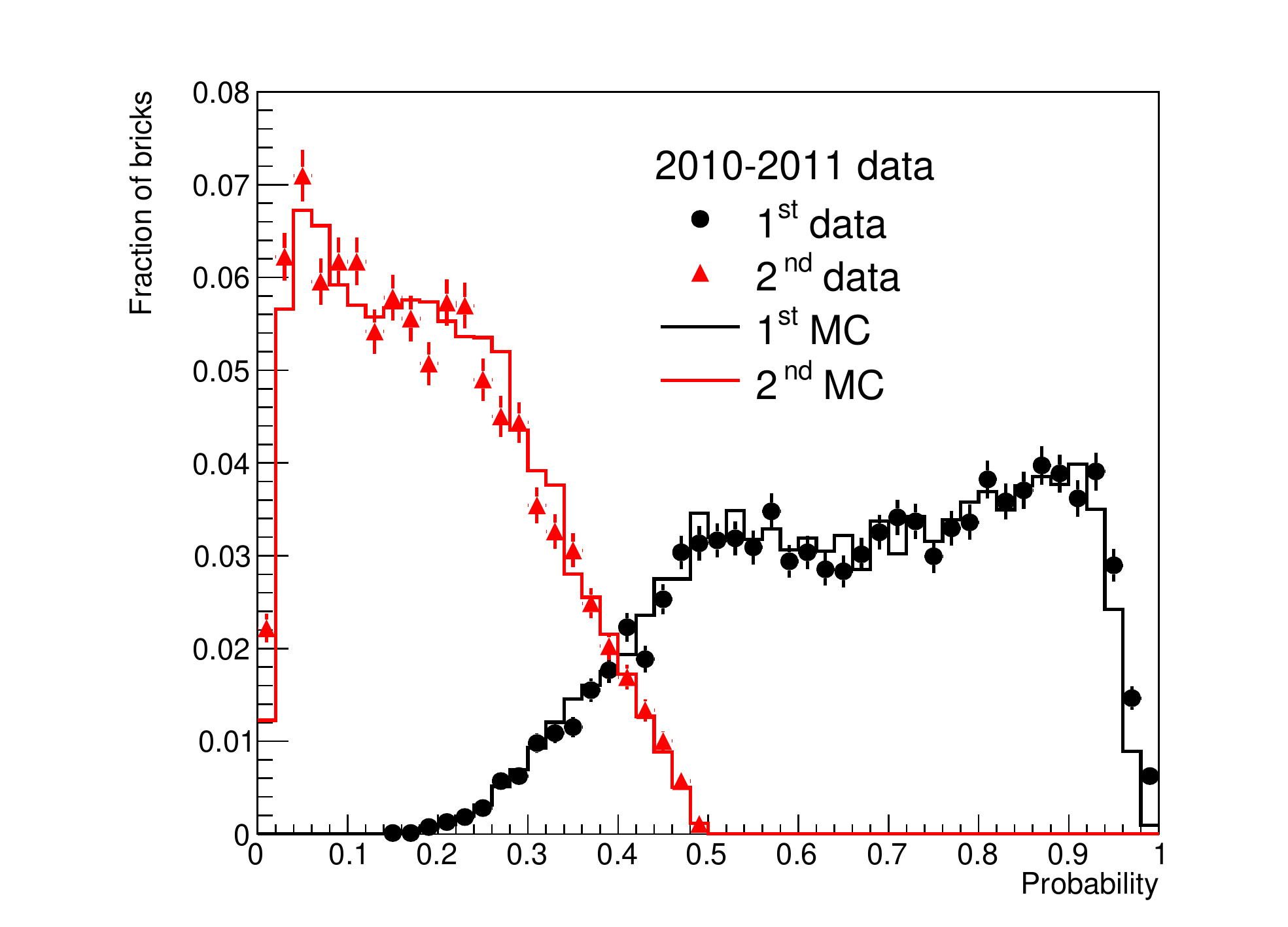}
\caption{Left: Shape comparison of the distributions of the
  reconstructed muon momentum ($p_\mu$) multiplied by the charge ($q$)
  for $1\mu$ events for 2008-2010 data (bullets) and MC (orange
  band). The pre-selection cut at $\pm~$15~GeV/$c$ is marked by the
  vertical lines. The bar height in the MC indicates a 10\% systematic
  error.  The dashed histogram represents the $\tau\to\mu$ channel MC
  simulation (shape only).  Right: normalized probability
  distributions for the 1$^{\rm{st}}$, 2$^{\rm{nd}}$ brick for the
  data of 2010 and 2011 (bullets) and the MC (histograms).}
\label{fig:pprob}
\end{figure}

In the $1\mu$-decay channel, with a branching ratio of \mbox{$(17.36
  \pm 0.05)$}\%, most events occur at low muon momentum. This is due
to the fact that the atmospheric-scale oscillation, at this baseline,
mainly affects the low-energy region. Moreover, a large fraction of
the $\tau$ momentum is transferred to the two final-state
neutrinos. This is illustrated in the left plot of
Fig.~\ref{fig:pprob} which also shows the distributions of the
reconstructed signed muon-momentum ($q \times p_\mu$) for
$1\mu$-events for data and Monte Carlo (MC). The shape of the data is
well described by MC, indicating that the pre-selection of events is
well understood.  The momentum cut at 15 GeV/$c$ (dashed lines)
results in a loss in the sample of muonic $\tau$ decays of only
{\color{black}{4\%}}. Instead, this cut reduces by
{\color{black}{33\%}} the amount of $\nu_\mu^{CC}$ events to be
analysed and, more important, by {\color{black}{28\%}} the size of the
charm sample.  The momentum cut at 15 GeV/$c$ is also applied to the
2008 and 2009 samples at a later step of the analysis chain in the
kinematical selection~\cite{proposal} (see, Sect.~\ref{subsec:kin}).

The right plot in Fig.~\ref{fig:pprob} shows the distributions of the
brick probability for $0\mu$ and $1\mu$ events of the 2010 and 2011
samples, separately for the first two bricks in the probability
ranking.  Again, the shape of the brick probability is well described
by MC. Details on the MC simulation will be given in
Sect.~\ref{sec:sim}.

The number of events selected at the level of electronic detectors
($N_{ED}$) is \NED~(listed for each year in the first row of
Tab.~\ref{tab:stats}). Part of this sample is discarded by the brick
finding algorithm based on the topology of the measured hits in the
TT.  The rejected sample is mainly accounted for by punch-through of
external interactions.  The numbers of events with a good brick
prediction, which are sent to the scanning laboratories ($N_{pred}$,
\NPRED~in total), are given in the 2$^{\rm{nd}}$ row in
Tab.~\ref{tab:stats}.

Selected bricks are routinely extracted by a robotic Brick Manipulator
System (BMS) capable of keeping up with the average weekly rate of
neutrino interactions and thus allowing for the emulsion-based
analysis to potentially proceed in parallel to the neutrino
interactions data taking.  Removed bricks with a negative CS result
are re-inserted in the target after replacement of the CS while
dismantled bricks are not replaced. Empty spaces are filled by a
rearrangement of bricks.  This approach aims at keeping the target
homogeneous and minimise the occurrence of events with irregularities
in the energy flow containment.  The position of each brick is
registered in a dynamic database to enable the event location.

\begin{figure}
\centering
\includegraphics[width=10cm]{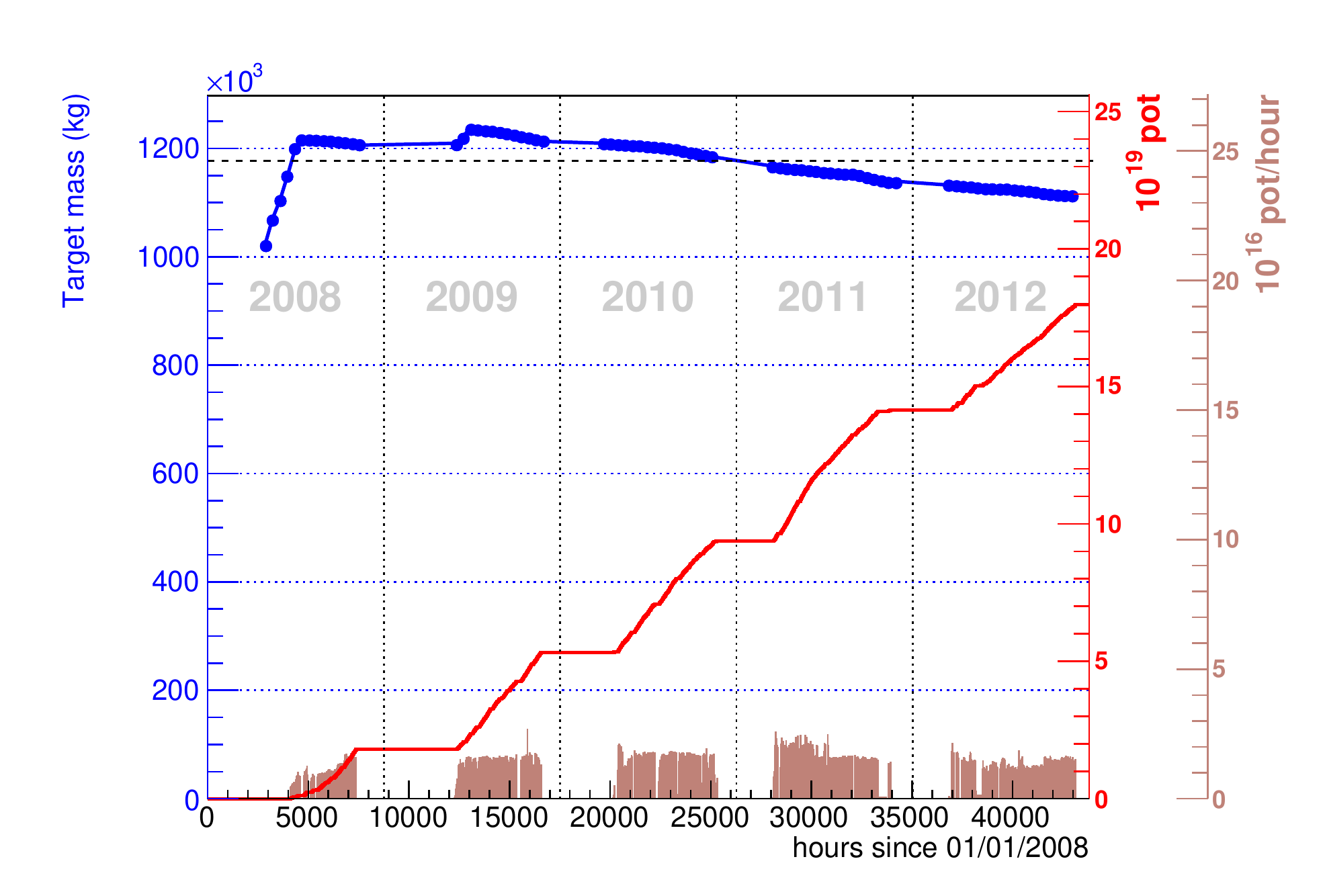}
\caption{Evolution of the detector target mass (blue bullets) and
the integrated number of pot (red line) over the total experiment running.
The filled histogram shows the pot in one hour wide bins. The dashed horizontal line
shows the pot-weighted average mass which amounts to 1.18 kt.}
\label{fig:massdecrease}
\end{figure}

The average number of bricks in the detector during each run is given
in the last column of Tab.~\ref{tab:samples}. The time evolution of
the target mass is shown in more detail in Fig.~\ref{fig:massdecrease}
(blue curve) together with the integrated number of pot (corrected for
the inefficiency of the DAQ system, red curve).  The loss in target
mass at the end of the running amounts to 8.5\% of its maximum value.

The pre-selected events ($N_{ED}$) represent a fraction of \FRACTEN~and 
\FRACELEVEN~of the 2010 and 2011 contained-event samples, respectively.

\begin{table}
\centering
{\small{
\begin{tabular}{c|cc|cc|cc|c|c}
\hline
\multirow{3}{*}{}
&
\multicolumn{2}{c|}{{\bf 2008}} & 
\multicolumn{2}{c|}{{\bf 2009}} & 
\multicolumn{2}{c|}{{\bf 2010}} &
\multicolumn{1}{c|}{{\bf 2011}} &
\multicolumn{1}{c}{{\bf 2008-2011}} \\
\hline
 & $0\mu$ & $1\mu$  
 & $0\mu$ & $1\mu$
 & $0\mu$ & $1\mu$, {\scriptsize{$p<15$ GeV/$c$}}
 & $0\mu$ & 
\multirow{2}{*}{all}\\
 & \multicolumn{2}{c|}{\scriptsize 2B}
 & \multicolumn{2}{c|}{\scriptsize 2B}
 & \multicolumn{2}{c|}{\scriptsize 1B}
 & {\scriptsize{1B}} &   \\
\hline
$N_{ED}$ 
& 552 & 1379  
& 1199 & 2806   
& 1420 & 2109 
& 1684 
&11149\\
$N_{pred}$ 
& 404 & 1328  
& 884 & 2696   
& 948 & 2014 
& 993  
& 9267\\
$N_{brick}$
& 152 & 848  
& 265 & 1597 
& 218 & 830 
& 225 & 4135\\ 
$N_{DS}$
& 146  & 819  
& 248 & 1554  
& 209 & 794 
& 199 & 3969\\
\hline
\end{tabular}
}}
\caption{Summary of the analysed data samples. 2B stands for two-brick
  analysis while 1B denotes the highest-probability brick analysis.
  $N_{ED}$ (Sect. \ref{sec:presel}) are the pre-selected events,
  $N_{pred}$ (Sect. \ref{sec:presel}) are the events with a brick prediction,
  $N_{brick}$ (Sect. \ref{subsec:vtxloc}) the events located in the
  bricks and $N_{DS}$ (Sect. \ref{sec:DS}) the number of events for
  which the decay-search procedure was applied. }
\label{tab:stats}
\end{table}

\subsection{Analysis of the Changeable Sheets}
\label{subsec:CS}
After the extraction of the selected brick from the detector, the
scanning of the CS doublets is performed in order to validate or
disprove the brick-finding result.  The pattern of tracks
reconstructed in the CS doublets can confirm the prediction of the
electronic detector, or act as veto and thus trigger the extraction of
neighbouring bricks.

CS doublets are inspected by automatic optical scanning
microscopes~\cite{ref14_d,uts_a,uts_b,uts_c,uts_d,ess_a,ess_b,ess_c,ess_d,ess_e}
in two specialised scanning stations at the \textsc{LNGS} laboratory
and at Nagoya University.  A rectangular shape is defined centered on
the prediction of the electronic detectors with a resulting average
scanning area of 20~cm$^2$ for $1\mu$ events and 35 cm$^2$ for $0\mu$
events, the pointing accuracy here being poorer due to the absence of
a muon track.

The tracking efficiency of the CS doublets has been measured in
neutrino data and test beam exposures~\cite{CSpaper,CSpaperJP}.  A
charged particle crossing the CS will produce 4 track segments called
micro-tracks.  The micro-track detection efficiency for minimum
ionising tracks is 95\%~\cite{CSpaperJP} with the automatic scanning
system.  The distributions of the residuals in position and angle
between the ED and CS tracks have a standard deviation of about 8~mm
and 15~mrad respectively.

The analysis and scanning strategy have been improved during the
years.  In the analysis of the 2008 and 2009 runs, a brick was sent to
development as soon as at least one CS track had a possible match with
some ED hits.  Subsequently the strategy \mbox{(``CS trigger'')} has
been refined for the 2010 and 2011 runs by requiring the presence of
at least:
\begin{itemize}
\item for $1\mu$ events, a CS track compatible with the ED muon track within 60 mrad
\item for 0$\mu$ events, a CS track matching an isolated ED track 
\item 2 or more CS tracks possibly converging towards a common origin in the brick.
\end{itemize}
If none of these conditions are fulfilled, the brick is put back in
the target with a new CS doublet and the next brick in the probability
map is extracted (in case of multi-brick analysis).  This method
allows saving scanning and analysis time and minimising the target
mass loss.  In case of a positive outcome, the brick is exposed to
cosmic rays in a dedicated pit in the surface \textsc{LNGS} laboratory
for 14 hours for high-precision film-to-film alignment, later to be
dismantled in the dark room where the emulsion films are developed.
The films are finally dispatched to the scanning laboratories of the
Collaboration for the \mbox{``vertex location''} and \mbox{``decay
  search''} analysis.

\subsection{Vertex location in the brick}
\label{subsec:vtxloc}
CS tracks are projected to the most downstream emulsion film in the
brick (through a distance of about 4.5~mm mainly filled with the
plastic and Aluminium film of the CS and brick boxes) where they are
searched for. The residuals in position are at the level of 50-60
$\mu$m such that the tracks are typically found within the predicted
microscope view (400 $\times$ 300 $\mu$m).  They are then followed
upstream film by film (``scan-back''\cite{scanback_a,scanback_b})
adjusting the predictions in angle and position at each step to cope
with multiple Coulomb scattering (MCS) in the lead plates and
measurement errors effects.

The scan-back procedure is stopped when no track candidate is found in
three consecutive films; the lead plate just upstream of the last
detected segment is defined as the candidate vertex plate.  At this
stage a volume is defined with a transverse area of 1 cm$^2$ for 5
films upstream and 10 films downstream of the stopping point (or less
if the stopping point lies too upstream or downstream in the brick)
and tracks within an angular acceptance $\tan \theta <0.6$ ($\theta$
being the angle of the track with the $z$ axis,
Fig.~\ref{fig:operadet}) are searched for in this volume (general
scan).

The scan-back procedure is modified for events in which the CS-trigger
is produced only by a shower-like topology (mainly due to
\mbox{$\gamma$ conversions} from $\pi^0$ decays). In such cases an
area of 1 cm$^2$ centered on the shower axis is analysed for 20 films
(corresponding to about 3.5~$X_0$) starting from the last plate.

All track segments collected in the scanned volumes are analysed by
offline algorithms which perform precise alignment, tracking and
vertexing.  The alignment of the films (with an accuracy of a few
$\mu$m) is assured by the cosmic ray tracks (Sect.~\ref{subsec:CS}).

In some cases the scan-back track is not associated to any other track
inside the volume due to a low-multiplicity event or because of the
angular acceptance limitation or tracking inefficiencies.  In this
condition the vertex is consider as detected. A further signature of
the neutrino interaction can then be obtained from the analysis of
nuclear fragments which might be visible in the forward and/or
backward hemisphere especially if the interaction happened close to
the emulsion layers.  Dedicated image analysis tools have been
developed~\cite{largeangle_a,largeangle_b} to detect highly ionising
large-angle tracks in the proximity of the stopping point.

Interactions are tagged as ``dead material'' when the extrapolated
vertex does not lie in the lead plates or in the films (i.e. it is in
the scintillator strips, in the brick supporting structure) and no
event-related tracks are found in the upstream-brick CS-doublet.

For the 2008-2011 analysed sample, the events located in the brick
lead plates ($N_{brick}$) amount to \NBRICK~(Tab.~\ref{tab:stats}).

\subsubsection{Data-MC comparison for the location efficiency}
An important check of the understanding of the complex task of event
location is done at this stage. We consider the data collected in the
year 2008 and 2009 and the two-brick analysis strategy.

For the definition of the initial control sample of $0\mu$ events
($N_{0\mu,ED}^\prime$) we adopt a stricter selection on the fiducial
volume and a minimal requirement on the energy released in the TT
\mbox{($E_{TT}>70$~MeV)}.  This allows getting the contamination of
external events to the level of 5\% in the $0\mu$ sample.  This
residual contamination has been statistically subtracted.  The
location efficiency
\mbox{$\epsilon_{loc}=N_{brick}/(N_{ED}^\prime(1-f_{DM})(1-f_{BQ}))$}
is then defined, $f_{DM}$ being the fraction of interactions in the
dead material which is estimated from the MC simulation (\mbox{$f_{DM}
  = 7.8$\%}) and $f_{BQ}$ being the fraction of films which could not
be analysed due to their bad quality ($f_{BQ}=6\%$).

\begin{figure}
\begin{center}
\includegraphics[width=9cm]{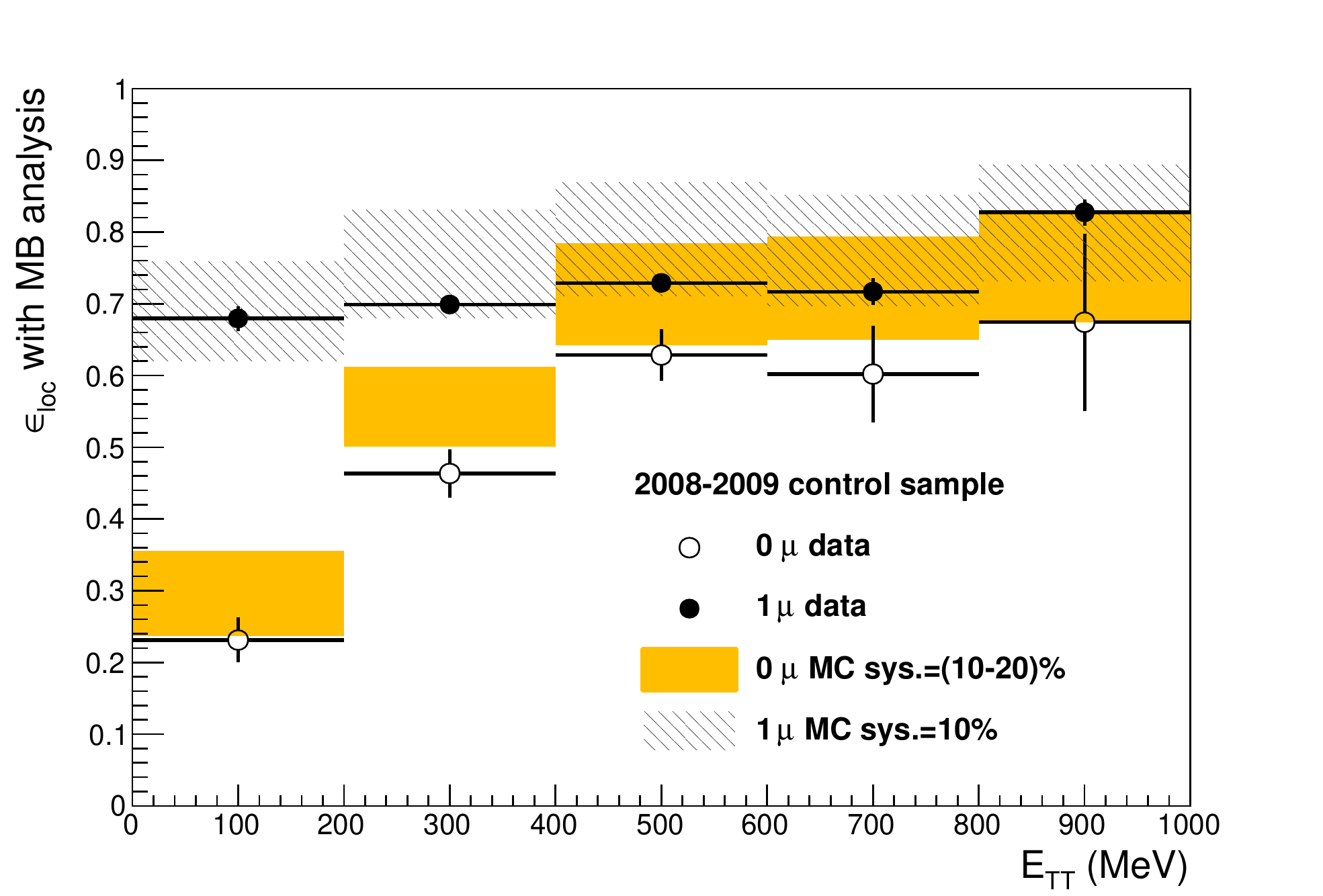}
\end{center}
\caption{Location efficiency with the two-brick analysis
  ($\epsilon_{loc}$) vs $E_{TT}$ in a control sample of data collected
  in 2008 and 2009 (bullets) compared with MC (histograms).  The
  comparison is done separately for $0\mu$ and $1\mu$ events.  The
  error bars represent the systematical and statistical
  uncertainties in the MC and in the data respectively.
\label{fig:effloc}
}
\end{figure}

The dependence of the location efficiency $\epsilon_{loc}$ on $E_{TT}$
is shown in Fig.~\ref{fig:effloc} for data and MC for $1\mu$ and for
$0\mu$ events.  The two quantities are highly correlated in the $0\mu$
sample where the hadronic activity plays a crucial role in the
location. The bands in the MC prediction are representing the
systematic uncertainty which is 10\% for the $1\mu$ sample and the
$0\mu$ sample with $E_{TT}>200$ MeV and 20\% for the $0\mu$ sample
with $E_{TT}<200$ MeV.  The systematic uncertainties account for
residual differences in the MC description of the data: the
implementation of the CS trigger and brick scanning strategy, the
definition of the dead material contribution. For the $0\mu$ sample an
additional uncertainty arises from the subtraction of the external
background component.

The agreement between data and MC is good in shape and normalisation
for the $1\mu$ sample while for the $0\mu$ sample the agreement is
good in shape, but data tends to be below the MC by about 15\%,
independently of $E_{TT}$.  It must be noted, however, that this
residual difference in the location efficiency in data and Monte Carlo
for the $0\mu$ sample has no effect on the predicted signal or
background events which is normalised to the number of localised
events in the data (see Sect.~\ref{sect:expect}).

\subsection{Decay search (topological selection)}
\label{sec:DS}
The decay search (DS) procedure is aimed at detecting the decay topologies
of $\tau$ leptons produced in $\nu_\tau^{CC}$ interactions once a
vertex has been identified in the volume scan data.  The decay is
defined as ``short'' if it happens in the same lead plate where the
neutrino interaction occurred or in the first downstream emulsion
layer and as ``long'' if it happens further downstream such that at
least one complete micro-track is produced by the $\tau$-lepton.
About 46\% of the $\tau$ decays are expected to be short (43\% in the
first lead plate and 3\% in the first emulsion layer) while the
remaining 54\% are long decays happening in the first plastic base
(11\%), in the second emulsion layer (2\%), in the second lead plate
(25\%) or further downstream (16\%).

Candidate daughter tracks from short-lived particles decays are
selected by requiring their impact parameter ($IP$) with respect to
the reconstructed neutrino interaction vertex to be larger than
10~$\mu$m if the depth in lead ($\lambda$) is lower than 500~$\mu$m or
by loosening this requirement to $IP > (5 + 0.01 \times
\lambda)$~$\mu$m for deeper vertices.  The next requirement is that
the momentum of the candidate daughter track measured from its MCS
($p_{mcs}$)~\cite{MCSpaper} is larger than 1~GeV/$c$.  If the number
of planes in the scanned volume is not sufficient to estimate
$p_{mcs}$ the angular spread $S_\theta$ of the available segments is
evaluated (see \cite{refDS} for a detailed definition).  If $S_\theta
> 15$~mrad in both views, the track is discarded, otherwise the MCS
measurement is extended to more plates.

A search is then performed for additional tracks (called extra-tracks)
which are not directly connected to the primary vertex: they must be
detected in at least 3 films and the distance along the $z$-axis
between the most upstream segment and the neutrino vertex ($\Delta
z=z_{up}-z_{vtx}$) is required to be positive and less than 3.6~mm. In
addition we require $IP<$~300~$\mu$m if $\Delta z<$~1~mm and
$IP<$~500~$\mu$m for $\Delta z>$~1~mm.

In order to deal with topologies in which the first-reconstructed
vertex is the decay vertex and thus downstream of the primary neutrino
vertex (e.g. in a $\tau\to 3h$ decay) the search for extra tracks is
also extended upstream, requiring $\Delta z>-2$~mm and
$IP<$~500~$\mu$m.  For the events located on the basis of a single
track (which could be due to a higher-multiplicity decay with tracks
outside of the scanning angular acceptance), extra-tracks are searched
for within a $\Delta z$ of $\pm$~3~mm and $IP<500$~$\mu$m.  In this
case the $z$ position of the vertex is assumed to be at the center of
the lead plate.

A search for charged short-lived parents is then applied to both
upstream and downstream extra-tracks by looking for tracks or single
segments with $IP<10$~$\mu$m and a distance of closest approach
$d_{ca}<$~20~$\mu$m with respect to the possible daughter tracks.

Finally the presence of a significant kink (larger than 20~mrad) is
checked on the muon track for $1\mu$ events and on all tracks for
$0\mu$ events.  The four most upstream segments of the track are used
to evaluate the ratio ($R$) between the maximum angular difference
between pairs of segments and the overall angular spread~\cite{refDS}.
If $R>5$ then a ``kink trigger'' is issued and further analysis is
performed.

All these operations are supplemented by manual eye-inspection at the
microscope in order to improve the information provided by automatic
scanning in terms of angular resolution (the automatic system accuracy
can sometimes be spoiled by the association of uncorrelated background
grains) and efficiency (by recovering the information for inefficient
planes). Eye-inspection is also employed to improve the purity by
discarding fake tracks (most important for large angles and
high-fog\footnote{AgBr crystals can be activated in the emulsion due
  to thermal or mechanical excitations resulting in the presence of
  randomly distributed grains which are referred to as ``fog''.  }
films), and furthermore allows excluding passing--through cosmic ray
tracks which can mimic extra-tracks due to inefficiencies of the
automatic scanning in the five upstream films.

The number of events with a completed decay-search ($N_{DS}$) for the
present analysis amounts to \NDS~(Tab.~\ref{tab:stats}). This number
differs from $N_{brick}$ by about \NOTDSONLOC~due to the events which
fall at the edges of the brick (either in the longitudinal plane or in
the transverse direction).

\subsubsection{The charm control sample}
\label{subsec:charm}
Given the similarity in mass and decay topologies, the detection of
charmed particles constitutes not only a background but also an
important tool to verify the understanding of the $\tau$ detection
efficiency up to the topological selection (decay-search) level.

The charm sample is selected using the same DS analysis as used for
the signal search with the exception that the kinematic selection of
$\nu_\tau$ candidate events (which will be described in the following
section) is not applied\footnote{For 1-prong decays a minimum kink angle 
of 20~mrad and a momentum of the daughter particle larger than 1~GeV/$c$ 
is also required.}.

\begin{figure}
\begin{center}
\includegraphics[width=6.5cm]{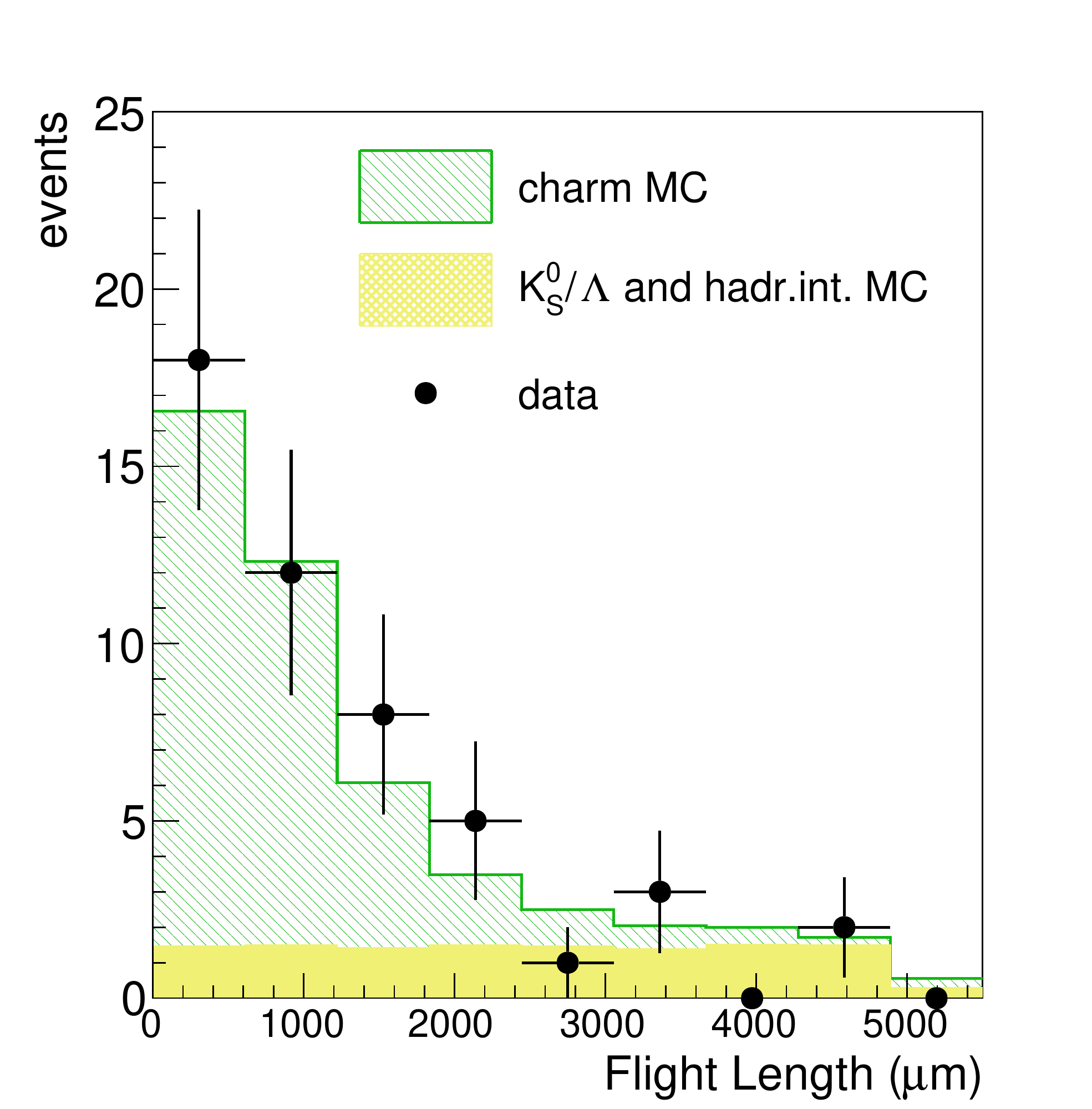}%
\includegraphics[width=6.5cm]{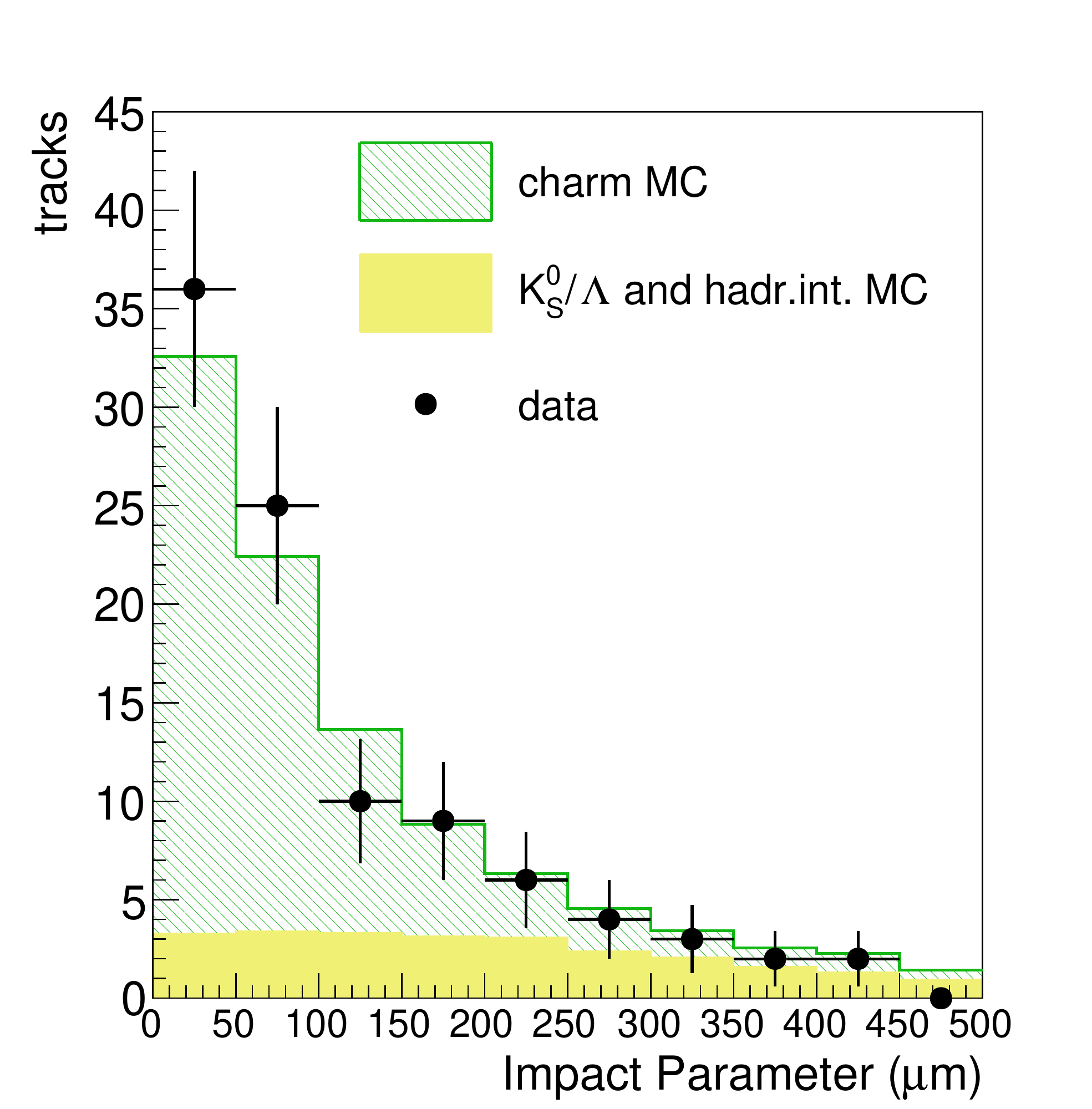}
\end{center}
\caption{Shape comparison of the distributions of the flight length
  (left) and the tracks' impact parameters with respect to the primary
  vertex (right) for the charm data sample of 50 events, described in
  Sect.~\ref{subsec:charm}, (bullets) and the MC simulation of
  $40 \pm 3$ expected charm events (green hatched histogram) and
  $14 \pm 3$ expected hadronic interactions and
  strange meson decays (yellow hatched
  histograms)\label{fig:figcharmkin}.
}
\end{figure}

The charm yield has been predicted using the latest analysis of the
\CHORUS~data~\cite{ref21} at the \WANF~neutrino beam ($\langle E_\nu
\rangle \simeq 27$~GeV) and re-weighting it for the different neutrino
energy spectrum at \OPERA~(see Sect.~\ref{subsec:charmbackground} for
more details).  The decay search efficiency is estimated to be
(58~$\pm$~8)\% for long charm decays and (18~$\pm$~2)\% for short
charm decays.  The main sources of background in the charm selection
are hadronic re-interactions (about $87\%$ of the total background)
and decays of $K^0_S$ or $\Lambda$.  In the analysed sample of events
from the 2008, 2009 and 2010 years having at least a muon tagged
3D-track, a total of ($40 \pm 3$) charm events and ($14 \pm 3$)
background events are expected while 50 charm candidate events are
observed in the data.  The distributions of the flight length of the
charm candidates and of the impact parameters of the secondary
particles with respect to the primary vertex are presented in
Fig.~\ref{fig:figcharmkin} for data and MC.  Not only the absolute
yields but also the shapes of the distributions are in very good
agreement, which indicates that the systematic error on the estimated
efficiency of the full analysis chain cannot exceed \charmuncertainty.
A more extensive discussion of the charm sample will be presented
in~\cite{nextcharmpaper}.

\subsection{Kinematic selection}
\label{subsec:kin}
Several kinematic quantities of the neutrino interaction are
accessible at brick-level via momentum reconstruction using MCS. The
energy of electrons and photons is also measured by employing
calorimetric techniques~\cite{EMrec,nuepaper}.  Kinematic criteria can
be defined to improve the signal-to-background ratio. To improve the
acceptance for electromagnetic showers and reduce the error on the
track momentum measurement the standard volume considered for the
location (Sect.~\ref{subsec:vtxloc}) is enlarged and tracks are
followed downstream, eventually in other bricks (Sect.~\ref{sec:TFD}).

After denoting the charged tracks emerging from the neutrino
interaction vertex or the decay vertex as ``primaries'' and
``daughters'' respectively, and the short-lived primary track as
``parent'', we define the following relevant variables:
\begin{itemize}
\item $z_{dec}$: the $z$-coordinate of the decay vertex with respect
  to the downstream face of the lead plate containing the primary
  vertex ($z_{dec}<44~\mu$m for short decays).
\item $p_{T}^{2ry}$: the transverse momentum of the daughter with
  respect to the parent direction, for 1-prong decays.
\item $p_T^{miss}$: the magnitude of the vectorial sum of the transverse 
  momenta of  primaries (except the parent) and daughters with respect to the
  neutrino beam direction.
\item $p^{2ry}$: the scalar sum of the momenta of the daughters.
\item $\theta_{kink}$: the average 3D angle between the parent and
  its daughters (kink angle).
\item $m$: the invariant mass of the daughters (calculated attributing
  the $\pi$ mass).
\item $m_{min}$: the minimal invariant mass~\cite{ref:mininvmass}.
\item $\phi_{lH}$: the angle between the parent and the vectorial sum
  of the other primaries calculated in the plane perpendicular to the
  \CNGS~axis.
\end{itemize}
In the calculation of $\phi_{lH}$, if the primary multiplicity
(including the $\tau$ track candidate) is larger than two, the
primaries are defined after removal of the track with the largest
difference in $\phi$ with respect to the $\tau$ candidate (TLD) unless
it is identified as a hadron with high probability by the track
follow-down procedure (Sect.~\ref{sec:TFD}). This is intended to
remove events where the TLD is actually a genuine muon which has not
been identified as such by the electronic detectors (for these events,
if the TLD would not be excluded the angle would be large and thus the
kinematic cut on this variable would not be effective).

The applied selections are summarised for each channel in
Tab.~\ref{tab:kinesel}.  The choice of cut parameters has been studied
and defined in~\cite{proposal} for the 1$h$, electron and muon
channels and in~\cite{tesiAntonia} for the 3$h$ channel.

\begin{table}
\centering
{\small{
\begin{tabular}{ccccc}
\hline
variable &$\tau\to 1h$&$\tau\to 3h$&$\tau\to\mu$&$\tau\to e$\\
\hline
lepton-tag &\multicolumn{4}{c}{No $\mu$ or $e$ at the primary vertex} \\
$z_{dec}$ ($\mu$m)&  $[44, 2600]$ & $<2600$ &   $[44, 2600]$  &   $<2600$ \\
$p_T^{miss}$ (GeV/$c$)&$< 1^\star$ &$< 1^\star$ & / & /\\
$\phi_{lH}$ (rad)& $> \pi/2^\star$ & $> \pi/2^\star$ & / & / \\
$p_{T}^{2ry}$ (GeV/$c$) & $>$~0.6(0.3)*& / & $>$~0.25 & $>$~0.1 \\
$p^{2ry}$ (GeV/$c$)& $>$~2 & $>$~3 & $>$~1 and $<$~15 & $>$~1 and $<$~15 \\
$\theta_{kink}$ (mrad)& $>~20$& $<~500$ & $>~20$ & $>~20$ \\
$m, m_{min}$ (GeV/$c^2$)& / & $>~0.5$ and $<~2$& / & / \\
\hline
\end{tabular}
}}
\caption{Kinematic selection. The meaning of the variables is defined
  in the text.  The cut on $p_{T}^{2ry}$ for the 1-prong hadronic
  decay is set at 0.3 GeV/$c$ in the presence of $\gamma$ particles
  associated to the decay vertex and to 0.6 otherwise.  Cuts marked
  with a $\star$ are not applied in the case of a QE event. Only long
  decays are considered for the $\tau\to\mu$ and $\tau\to h$ channels
  due to a large background component in short decays from charmed
  particles and hadronic re-interactions respectively.
\label{tab:kinesel}}
\end{table}

\subsection{Track follow-down}
\label{sec:TFD}
The track kinematics of interesting events fulfilling the selection
are further studied in neighbouring bricks.  Primary tracks are
followed (with an angular acceptance extending up to $\tan \theta=1$)
until either a stopping point, an interaction or a muon decay topology
is found (track follow-down, TFD).  Thus, by the study of
momentum-range correlations, track length, energy loss in proximity of
the stopping point, and (if possible) the tagging of interactions or
muon decays, a muon/hadron separation exceeding the limitations of the
electronic detector reconstruction is obtained.

This procedure results in a significant reduction of the background in
all channels from $\nu_\mu^{CC}$ charm production where the primary
$\mu$ is not identified in the ED. The TFD technique also provides an
important reduction of the background in the $\tau \to \mu$ channel
from $\nu_\mu^{CC}$ events in which the primary $\mu$ is wrongly
associated to a secondary vertex with a kink topology from a hadronic
re-interaction. This also holds for $\nu_\mu^{NC}$ events with a fake
muon. The improved muon/hadron separation provided by TFD is also
effective in reducing the background in the hadronic channels due to
hadron re-interactions in $\nu_\mu^{CC}$ events where the electronic
detectors alone do not allow the primary $\mu$ to be identified
unambiguously.

\begin{figure}
\begin{center}
\includegraphics[width=9cm]{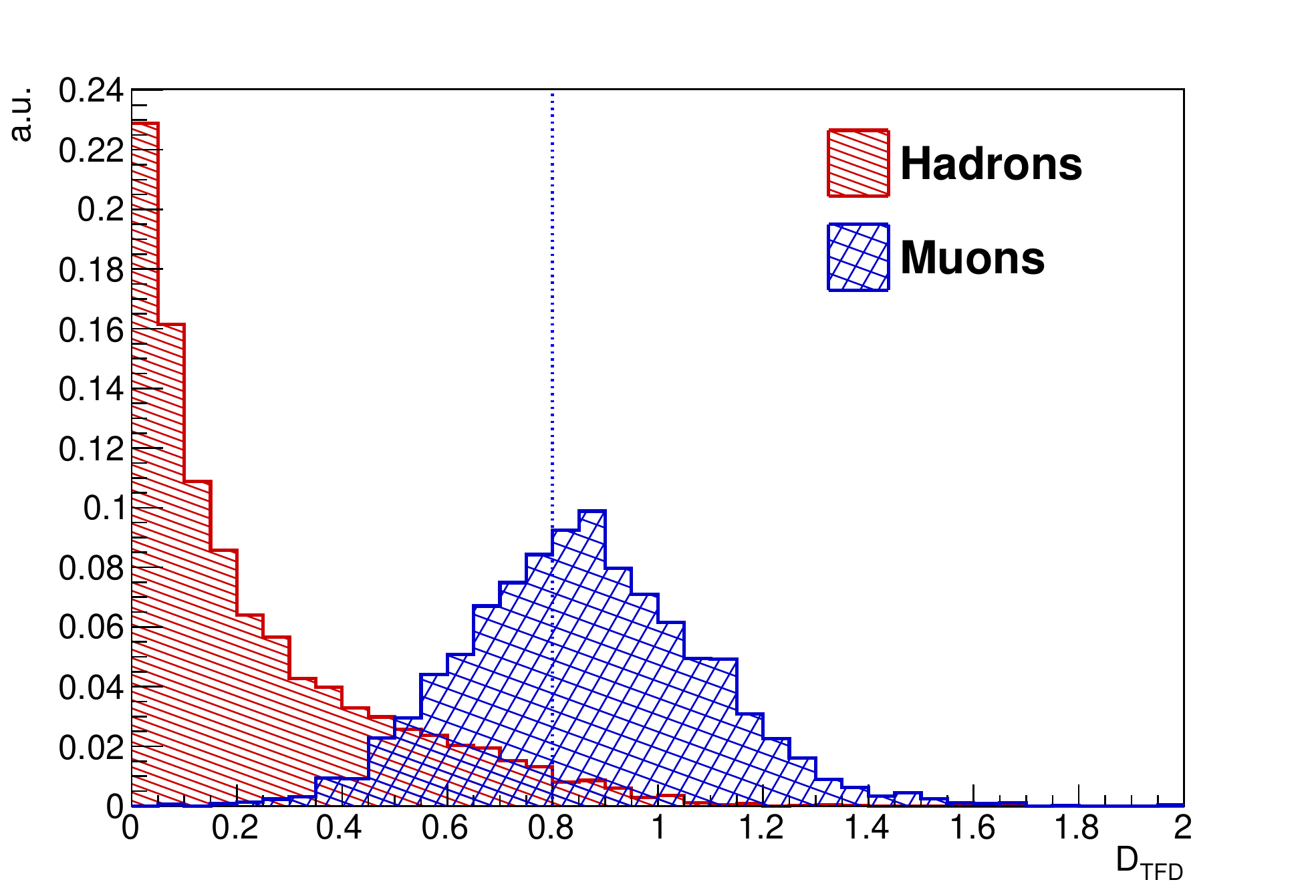}
\end{center}
\caption{MC distributions of the $D_{TFD}$ variable
  (Sect.~\ref{sec:TFD}) used to combine the information on momentum
  and range to separate muons from hadrons. The red (gray) histograms
  refers to genuine muons (hadrons). The vertical line denotes the
  used cut.  }

\label{fig:TFD}
\end{figure}

Momentum-range correlations are characterised by a discriminating
variable defined as: $D_{TFD} = \frac{L}{R(p)}\frac{\rho}{\langle
  \rho\rangle}$ where $L$ is the track length, $R(p)$ is the range in
lead of a muon with momentum $p$, $\langle \rho \rangle$ is the
average density along the path and $\rho$ is the lead density.  The MC
distributions of $D_{TFD}$ for hadrons and muons are reported in
Fig.~\ref{fig:TFD}.  If $D_{TFD} > 0.8$ the track is classified as a
muon. Among all the criteria used to separate muons from hadrons,
momentum-range correlations and energy loss close to the stopping
point, are those having a lower purity in the muon-tagging.  For this
reason, TLD tracks which are classified by the TFD as hadrons only by
one of the above criteria are not included in the calculation of
$\phi_{lH}$.

\section{Monte Carlo simulation of signal and backgrounds}
\label{sec:sim}

Compared to the results published in \cite{ref2ndoscpaper} several
improvements have been adopted in the simulation of the
detector. Efficiencies and backgrounds are based on a new software
framework which combines the information of the electronic and ECC
detectors.  All particle trajectories are digitised at the micro-track
level in a volume consisting of 3$\times$3$\times$3 bricks, with the
brick containing the neutrino interaction at the center. For dedicated
studies (i.e. for the TFD simulation), also larger volumes have been
considered.  The efficiency and the resolution (in angle and position)
of the scanning microscopes are simulated using parametrisations
obtained from real data.  The framework allows reproducing the
analysis flow from the prioritisation based on the brick probability,
the CS-trigger, the CS-to-brick connection, the scan-back, the
alignment and the vertexing with full access to the reconstruction
parameters.  Using this tool the decay search procedure has been
included in the efficiency evaluation. Furthermore the
detection efficiencies have been re-evaluated for all channels and the
simulation of the brick extraction strategies for the different years
of data taking has also been updated.

\subsection{Expected signal event rates}
\label{sect:expect}

The neutrino fluxes used in the calculation of the expected signal
rates are based on a \FLUKA~\cite{fluka} simulation of the
\CNGS~beam-line (2005 revision~\cite{CNGSflux}). The neutrino
interactions in the detector are generated using the NEGN generator
after a tuning of the parameters based on the \mbox{high-statistics}
data sample of the NOMAD experiment~\cite{NOMADtuning} .  The
energy-dependence of the $\nu_\tau$ cross section that has been used
is the default implementantion contained in the \GENIE~v2.6 simulation
program \cite{GENIEcit}.  The $\nu_\mu \to \nu_\tau$ oscillation
probability was evaluated using \mbox{\color{black}{$\Delta m^2_{23} =
    2.32 \times 10^{-3}$}~eV$^{2}$} \cite{PDGdm2} and \mbox{$\sin^2
  2\theta_{23}=1$}.  With these ingredients, the expected rate of
$\nu_\tau^{CC}$ on lead at true level is equal to
\mbox{{\color{black}{$3.32$}} events/(10$^{19}$ pot kt)}.  Taking into
account the time evolution of the pot delivery rate and the effective
mass of the detector throughout the running period
(Fig.~\ref{fig:massdecrease}) a total number of {\color{black}{66.4}}
$\nu_\tau$ CC interactions in lead are expected (with the non-DIS
component being about one third \cite{GENIEcit} of the total).

The expected rate of $\nu_\tau^{CC}$ events in the $0\mu$ sample can
be expressed in terms of the measured event rate in the same category
($n^{0\mu}$ from both $\nu_\mu^{CC}$ and $\nu_\mu^{NC}$) as:
\begin{equation}
n^{0\mu}(\nu_\tau^{CC}) = \frac{n^{0\mu}}{N(\nu_\mu^{CC})}\frac{\langle \epsilon^{0\mu}(\nu_\tau^{CC})\rangle}{\langle \epsilon^{0\mu}(\nu_\mu^{CC}) \rangle+\alpha \langle \epsilon^{0\mu}(\nu_\mu^{NC}) \rangle}
\label{eq0}
\end{equation}
denoting with $\langle \epsilon^r(k)\rangle$ the efficiency for
reconstructing the process $k$ ($\nu_\mu^{CC}$, $\nu_\mu^{NC}$) in the
reconstructed category $r$ ($0\mu$, $1\mu$) after convolution with the
\CNGS~flux $\phi$; $N(\nu_\mu^{CC})$ being the $\nu_\mu^{CC}$
interactions at true level and
$\alpha=\sigma(\nu_\mu^{NC})/\sigma(\nu_\mu^{CC})~\otimes~\phi$.  In
the present analysis the actual number of $n^{0\mu}$ is taken from the
data while other quantities are MC driven.  The expected numbers
obtained with the above-mentioned procedure are insensitive to
systematic effects on the efficiencies up to the location level being
common to $\nu_\tau$ and $\nu_\mu$ events.  The same approach is
followed in the prediction of the $\nu_\tau$ component in the $1\mu$
sample.  In this particular case, in addition to the standard $1\mu$
classification (Sect. \ref{sec:evsel}), the presence of a muon tagged
3D track is also required due to the need of matching this ED track to
the ECC secondary $\mu$ candidate track for the sake of background
reduction.

Finally it must be noted that in the present approach the signal
efficiency is not taking into account possible migrations between
different channels and could therefore be slightly underestimated.

\subsection{Expected background event rates}

Three sources of backgrounds are giving significant contributions to
the final sample: charmed particles decays, hadronic interactions and
large-angle muon scattering (LAS).  The last one only affects the
$\tau\to\mu$ channel while the other classes differently affect all
decay channels. These backgrounds have been discussed extensively
in~\cite{proposal}. The basic ingredients for the more recent
evaluation of these components will be discussed in the following.

\subsubsection{Charmed particle decays.}
\label{subsec:charmbackground}

The most effective tool for the rejection of the background from
charmed particles is an efficient identification of the primary muon
in $\nu_\mu^{CC}$ interactions. The fraction of $\nu_\mu^{CC}$ events
having an associated charm quark at the \CNGS~energies is estimated
from the \CHORUS~measurement \cite{ref21} as
$(\sigma(\nu_\mu^{CC}+c)/\sigma(\nu_\mu^{CC}))\otimes \phi=(4.38 \pm
0.26)\%$.  The tagging of the primary muon is achieved first, at the
level of the ED, via the reconstruction of penetrating 3D-tracks and
the vetoing of events involving many TT planes (to complement the
tracking) as described in Sect.~\ref{sec:evsel}.  Requiring an event
to be classified as $0\mu$ reduces the yield down to 6\% of the
initial charm sample. At the emulsion detector level, the TFD
procedure gives a further suppression by 60\%.  Taking into account
the effects of the full analysis chain and considering only the HPB,
an overall suppression of \cOnehsuppr is obtained in the $1h$ channel
and \cThreehsuppr in the $3h$ channel.  The two-brick analysis gives
an increase by a factor 20-25\% resulting in an expectation of
\COnehfinal events and \CThreehfinal events in the analysed sample for
the $1h$ and $3h$ channels respectively~(Fig.~\ref{fig:sigback},
green-coloured histogram and Tab.~\ref{tab:sigback}).

The background from charmed particles in the $\tau\to e$ decay channel
(where, unlike in the muon case, the positive charge of the decay
electron cannot be measured) gives a contribution in the analysed
sample of \CElecfinal events. This is the only significant background
in this channel.

The background from charmed particles in the $\tau\to\mu$ decay
channel is relatively small since it only arises in two cases i.e. 1)
when the primary $\mu$ is not identified, a muonic decay occurs
($f(C^+ \to \mu^+ h^0) = (5.3 \pm 2.8)\%$), and the positive charge of
the secondary $\mu$ is not measured or mis-identified; 2) the
secondary $\mu$ is not identified and the primary $\mu$ is wrongly
matched to the decay daughter.  This results in a background
contribution of \CMuonfinal events (Fig.~\ref{fig:sigback},
green-coloured histogram and Tab.~\ref{tab:sigback}).

Finally, the double charm production in NC interactions process has to
be considered. The yield has been measured by the
\CHORUS~collaboration to be $\sigma(c\bar{c}\nu)/\sigma_{NC}^{DIS} =
(3.62^{+2.9}_{-2.4}\rm{(stat.)}\pm {0.54}\rm{(sys.)}) \times
10^{-3}$~\cite{CHORUSDC}.  This component cannot profit of the
suppression given by the primary muon identification but, on the other
hand, it can be removed by measuring the presence of the primary and
secondary vertices. This contribution, which is expected to be a
second order correction, is not included in the present estimates.

The uncertainty on the charm background component is estimated within
the {\color{black}{20\%}} level based on the measured sample of charm
events (Sect.~\ref{subsec:charm}).

\subsubsection{Hadronic re-interactions.}
This background has been estimated by giving hadronic tracks from
located $0\mu$ events (from an initial sample of 9$ \times $10$^6$
$\nu_\mu^{NC}$ interactions) as an input to a \FLUKA-based MC
simulation.  With respect to a similar analysis reported
in~\cite{ref2ndoscpaper} this approach implements the \FLUKA~hadronic
models to fully reconstructed events thus including the biases
introduced by the selection chain, up to the event-location level.

The fraction of located $0\mu$ events with at least one track
mimicking the topology of a single-prong long decay and
\mbox{$\theta_{kink}> 20$~mrad} is found to be
{\color{black}{$2.1$\%}} while the fraction with at least a track
producing three visible prongs in the scanning acceptance is
{\color{black}{$0.18\%$}} (short and long).  The kinematic selection
(in particular the $p^{2ry}$ and $p_T^{2ry}$ cuts) strongly reduces
the ``topological'' sample to
{\color{black}{$(0.27~\pm~0.02(\rm{stat.}))\%$}} ($1h$) and
{\color{black}{$(1.2~\pm~0.1(\rm{stat.}))\%$}} ($3h$), due to the
typical low transverse momentum of the secondary products.  At this
stage an additional reduction of {\color{black}{$30\%$}} is obtained
by requiring the absence of nuclear fragments (either in the backward
or forward hemisphere) or minimum ionising particles up to
$\tan\theta=3$.  Finally the estimated rate of background events
amounts to {\color{black}{$(3.9~\pm~0.2(\rm{stat.}))\times 10^{-5}$}}
of the located $0\mu$ events for the $1h$ channel and
{\color{black}{$(1.5~\pm~0.2(\rm{stat.}))\times 10^{-5}$}} for the
$3h$ channel. The expected interaction rate per unit length in the
plastic base is about $\color{black}{4.5}$ times smaller than in the
lead plate due to the combined effect of the density and the mean
atomic number.

Several data-driven checks of the \FLUKA~description of hadronic
interactions in the \OPERA~bricks were
performed~\cite{ref2ndoscpaper}.  In general a good agreement between
data and simulation is observed for different data-sets: a sample of
hadronic interactions of pions of 2, 4 and 10 GeV/$c$ momenta from a
\CERN-based test beam; a sample of hadronic tracks from \CNGS~neutrino
interactions measured in the emulsions (total length 19~m) and a
sample of hadronic nuclear fragments from test-beam pion interactions
for which the yields and angular distributions have been studied. With
respect to~\cite{ref2ndoscpaper}, analyses based on larger data
samples have been developed~\cite{TohoArticle,tesiToho,tesiRescigno,tesiStellacci}
allowing estimating an accuracy of the predictions for the hadronic
background at the {\color{black}{30\%}} level.  The uncertainty in the
rate of high-angle nuclear fragments emitted in hadronic
re-interactions is estimated in~\cite{TohoArticle} at the 10\%
level. In conclusion, considering the 1$h$ and 3$h$ channels together,
in the analysed sample a total of \HOneThreefinal events from hadronic
re-interactions is expected (Fig.~\ref{fig:sigback}, yellow-coloured
histogram and Tab.~\ref{tab:sigback}).

\subsubsection{Large-angle muon scattering.}
The occurrence of large-angle scattering of muons in thin
(${\mathcal{O}}(0.1)X_0$) lead plates is, at present, not well
constrained by measurements.  Upper limits from extrapolations of
measurements on copper or nuclear emulsions have been reported
in~\cite{proposal}: the rates in the signal region
\mbox{$\theta_{kink}>20$~mrad} and \mbox{$p_T^{\mu}>0.25$~GeV/$c$} for
muon tracks with a realistic angular and momentum spectrum, are, at
\mbox{90\% C.L.}, below $2.3 \times 10^{-5}$ and $4.1 \times 10^{-5}$,
respectively.  A \mbox{\textsc{GEANT}~3.21}~\cite{G3} based simulation,
modified to take into account data on the nuclear form factor of lead
and inelastic interactions, predicts a rate of $2 \times 10^{-6}$.
More experimental activities to determine this process are in
progress. In the present work the same contribution used for the
experiment proposal~\cite{proposal} ($1 \times 10^{-5}$) is assumed
corresponding to \LASfinal events in the analysed sample
(Fig.~\ref{fig:sigback}, blue-coloured histogram and
Tab.~\ref{tab:sigback}).

\subsubsection{Summary of the expected event numbers.}
\begin{figure}
\begin{center}
\includegraphics[width=14cm,angle=0]{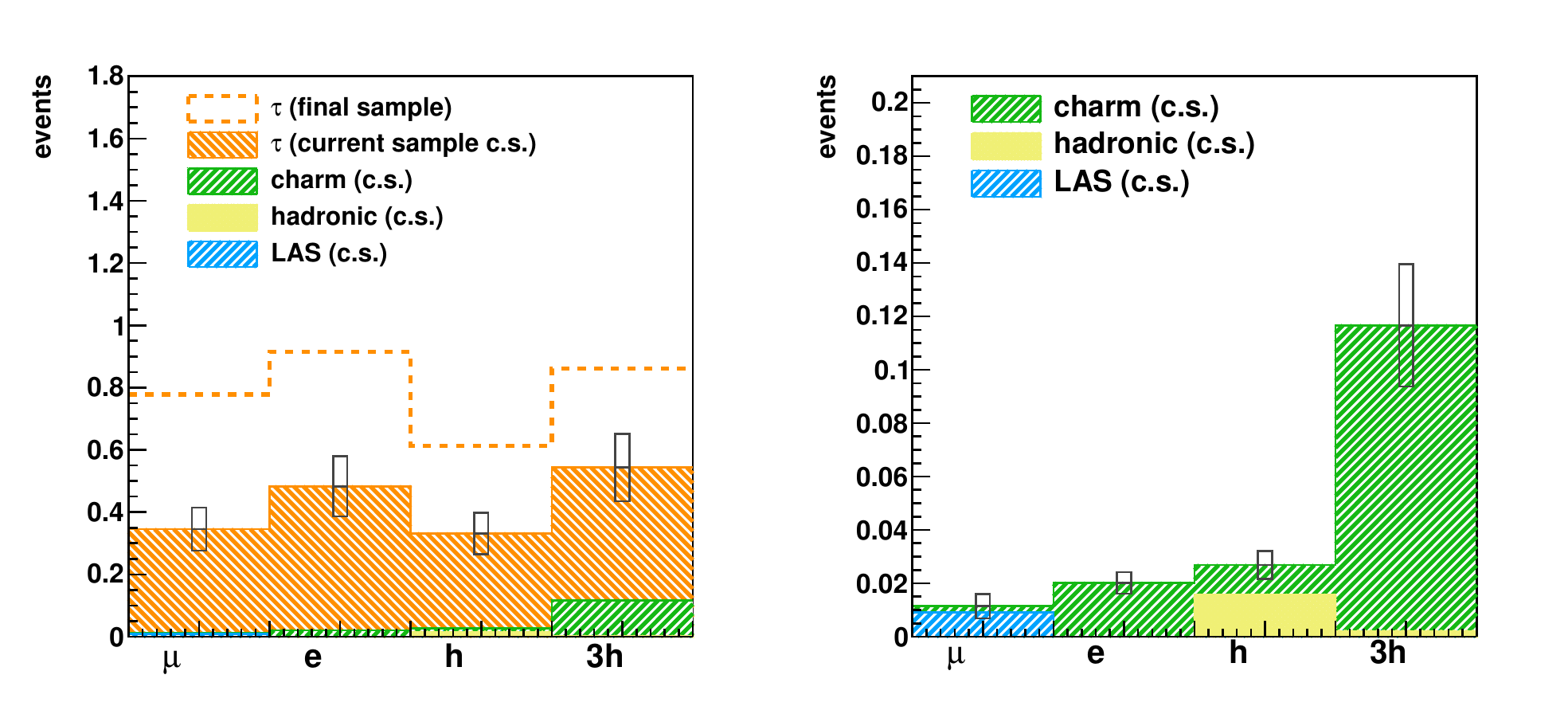}
\end{center}
\caption{Signal and backgrounds (left panel) and backgrounds only
  (right panel) expectations by channel for the sample considered in
  this paper (``c.s.'', filled histograms). Black rectangles show the
  estimated uncertainty. The dashed line in the left plot shows
  the signal expectation for the two-brick analysis of the complete
  data sample collected in the year 2008-2012.}
\label{fig:sigback}
\end{figure}

\begin{table}
{\footnotesize{
{\color{black}{
\begin{tabular}{cccccc}
\hline
& Signal  events & All  &Charm &LAS & Hadronic\\
& {\tiny{$\Delta m^2_{23} = 2.32 \rm{m(eV}^2)$}}& backgrounds & background & background &  background\\
\hline
$\tau \to h$        & $0.31~\pm~0.06$& $0.027~\pm~0.005$ &$ 0.011~\pm~0.002$& /  &  $0.016~\pm~0.005$\\
$\tau \to 3h$    & $0.43~\pm~0.09$ & $0.12~\pm~0.02$ & $0.11~\pm~0.02$ & /  &  $0.0021~\pm~0.0006$\\
$\tau\to\mu$  & $0.33~\pm~0.07$ & $0.012~\pm~0.005$ & $0.0023~\pm~0.0004$ & $0.009~\pm~0.005$ & /  \\
$\tau\to e$       & $0.46~\pm~0.09$ & $0.020~\pm~0.004$ & $0.020~\pm~ 0.004$ & /  & /  \\
\hline
all  & $1.53~\pm~0.16$ & $0.175~\pm~0.024$ & $0.15~\pm~0.02$  & $0.009~\pm~0.005$ & $0.018~\pm~0.005$\\
\hline
\end{tabular}
}}
}}
\caption{Signal and backgrounds expectations for the analysed
  sample. The numbers correspond to the dashed histograms in
  Fig.~\ref{fig:sigback}}
\label{tab:sigback}
\end{table}

The estimated signal and background events for the sample considered
in the present analysis are summarised in a graphical form in
Fig.~\ref{fig:sigback}, for each decay channel and in tabular form in
Tab.~\ref{tab:sigback}.  The background from charmed particles is the
dominant one in the $\tau\to e$ and $\tau\to 3h$ channels while LAS in
the $\tau\to \mu$ channel and hadronic re-interactions in the $\tau
\to 1h$ channel are the largest backgrounds.  The systematic
uncertainty on the signal and on the charm background is estimated to
be {\color{black}{20\%}} (Sect.~\ref{subsec:charm}) and the one on the
hadronic background to be {\color{black}{30\%}}. The LAS background
uncertainty is assumed to be of the order of {\color{black}{50\%}}
(grey bars in Fig.~\ref{fig:sigback}).

The expectation for the signal events using the two-brick analysis in
the final data sample (2008-2012) is of \ALLfinal observed $\nu_\tau$
events (dashed line in Fig.~\ref{fig:sigback}, left).  This number is
calculated using $\Delta m^2_{23} = 2.32 \times 10^{-3} \rm{eV}^2$ as
a central value\footnote{The number of oscillated events has a
  quadratic dependence on $\Delta m^2_{23}$.}, the accumulated
statistics of \FinalPOT~pot and a dynamic target mass
(Fig.~\ref{fig:massdecrease}) corrected for the dead material and the
fraction of bad-quality films (Sect.~\ref{subsec:vtxloc}).
Furthermore this estimate is based on a realistic simulation of all
the steps of the analysis chain and in particular the decay search
phase (Sect. \ref{sec:DS}) which had not been simulated in full detail
in previous analyses.

The current estimate does not include some factors which are expected
to increase the efficiency. The extension of the analysis to the
3$^{\rm{rd}}$ and 4$^{\rm{th}}$ bricks in the probability map is
foreseen. Smaller effects due to the migration of $\nu_\tau$ events
from one channel to another are not taken into account at
present. These effects increase only the expected number of $\nu_\tau$
events, while keeping the background at the same level.  Finally an
optimisation of the selection is ongoing within the new simulation
framework with the goal of maximising the expected sensitivity.

\section{Analysed sample and results}
\label{sec:res}

The number of events from the present analysed sample surviving the
complete selection chain described above amounts to two
$\tau$-candidates (hereafter called A and B). Given that the candidate
which occurred first (A, $\tau \to 1h$) has already been extensively
documented in ~\cite{ref1sttau,ref2ndoscpaper}, this paper will focus
on the description of the second candidate.

\subsection{Description of the second $\nu_\tau$ candidate event.}
\label{sec:tau2desc}
\begin{figure}
\centering
\begin{center}
\includegraphics[width=14cm]{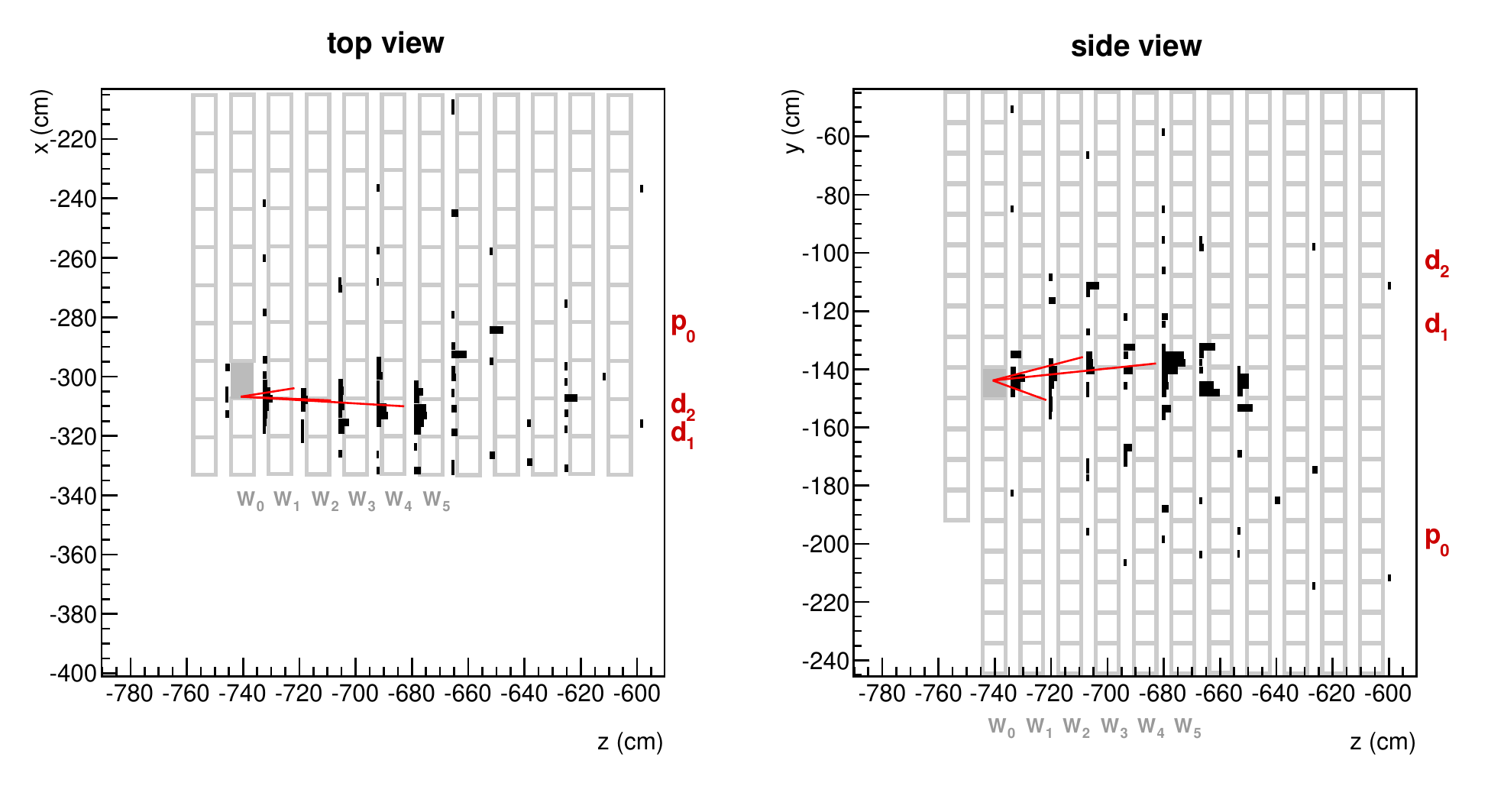}
\end{center}
\caption{Event display of $\nu_\tau$ candidate event B: Target Tracker hits
  with brick tracks super-imposed. The left panel is the top-view, the
  right panel the side-view. The position of the brick containing the
  neutrino interaction is highlighted.
\label{fig:fig_TTtau2}}
\end{figure}

This neutrino interaction occurred on \mbox{23 April 2011} at 7.15 UTC
time.  The pattern of hit scintillator strips in the TT is shown in
Fig.~\ref{fig:fig_TTtau2}. The event, of which the estimated hadronic
energy is ($22.0\pm 6.2$)~GeV, is classified as $0\mu$ and the bulk of
the activity in the electronic detector is contained within about 6 to
8 brick walls (more than 60 $X_0$ and about 2.5 pion interaction
lengths).  The interaction took place in the target of the upstream
Super Module and lies well within the brick-filled target region.  The
neutrino vertex brick (having a probability of {\color{black}{63\%}})
was located in the longitudinal direction~($z$), in the second most
upstream brick layer (called $W_0$), in the horizontal
direction~($x$), in the 3$^{\rm{rd}}$ brick layer from the closest
side (left side looking towards the \CNGS) and, in the vertical
direction~($y$), in the 19$^{\rm{th}}$ brick layer from the bottom. In
Fig.~\ref{fig:fig_TTtau2} the position of the brick containing the
interaction is highlighted. A linear extrapolation of the vertex
tracks found in the emulsions (described in the following) is also
shown to illustrate the matching with hits in the scintillators.
The event emulsion data has been independently measured with the
European and Japanese scanning systems with consistent results. The
average values are considered in the following.
The interaction is well inside the brick in the longitudinal
coordinate lying slightly upstream of the center of the brick
(22$^{\rm{nd}}$ lead plate in increasing $z$ order) while in the
transverse plane it lies at 6.8~mm and 41.4~mm from the closest sides
in $x$ and $y$.

\begin{figure}
\begin{center}
\includegraphics[width=10cm]{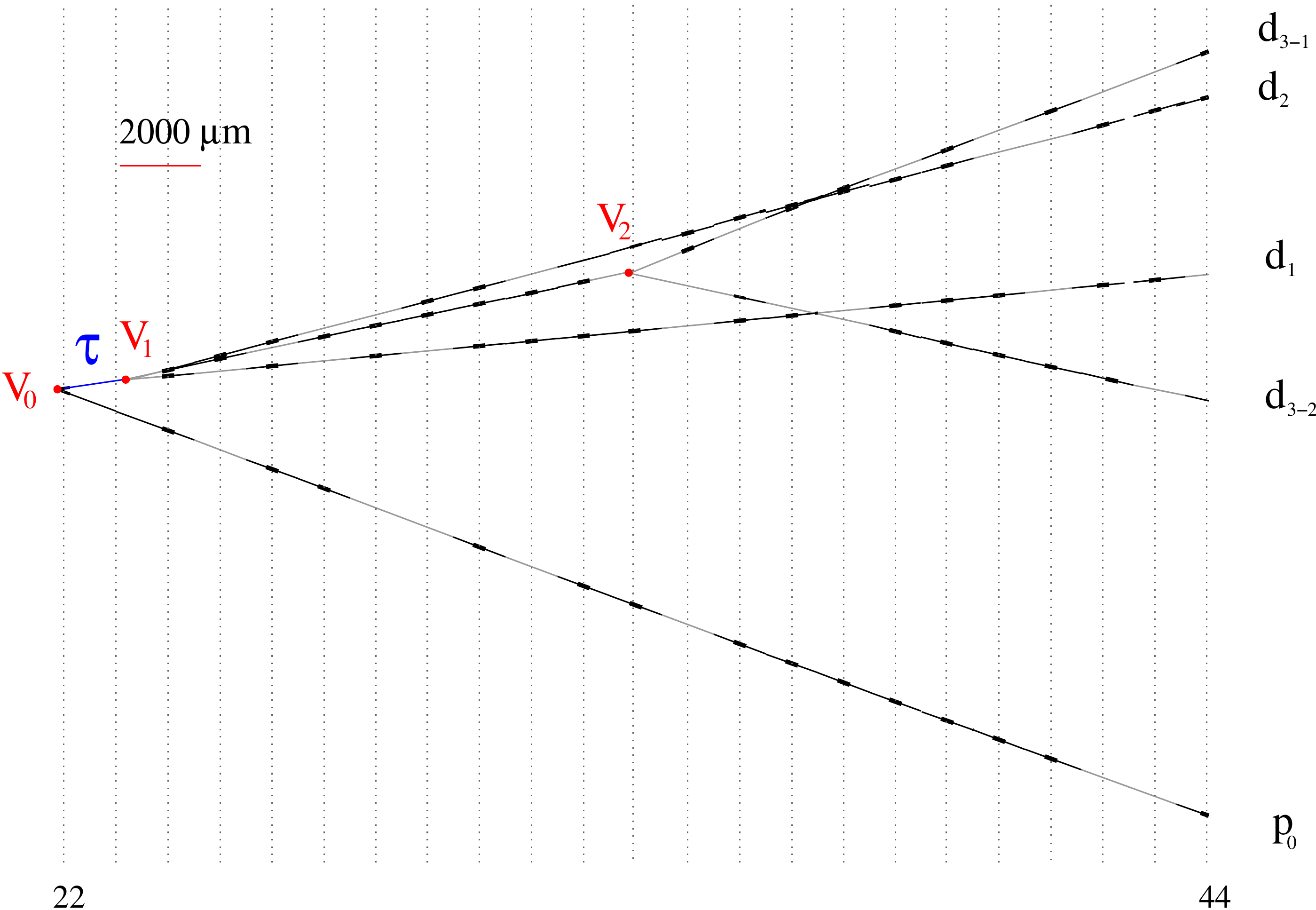}
\end{center}
\caption{Event display of $\nu_\tau$ candidate event B: reconstruction
  in the brick (side-view). Vertical lines indicate the position of
  the middle-point of emulsion films 22 to 44, numbering them in order
  of increasing $z$ from 1 to 57. The pitch is 1.3 mm. The black
  segments represent reconstructed base-tracks while the gray lines
  are the result of the track fit.
\label{fig:fig_tau2a}}
\end{figure}

The topology of the primary vertex ($V_0$ in Fig.~\ref{fig:fig_tau2a})
consists of two tracks, the $\tau$ lepton candidate and another track
(called $p_0$).  The distance of closest approach of the $p_0$ and
$\tau$ tracks equals to 0.2~$\mu$m and the vertex lies close to the
downstream emulsion film, at a depth in lead of only 120~$\mu$m.  A
\mbox{forward-going} nuclear fragment associated to the primary vertex
has also been detected at a large angle with slopes of: $(1.15,
-0.28)$\footnote{Slopes are given as tangents of the projected angles
  after correction for the 58 mrad vertical tilt of the beam.  }.

\begin{figure}
\begin{center}
\includegraphics[width=10cm]{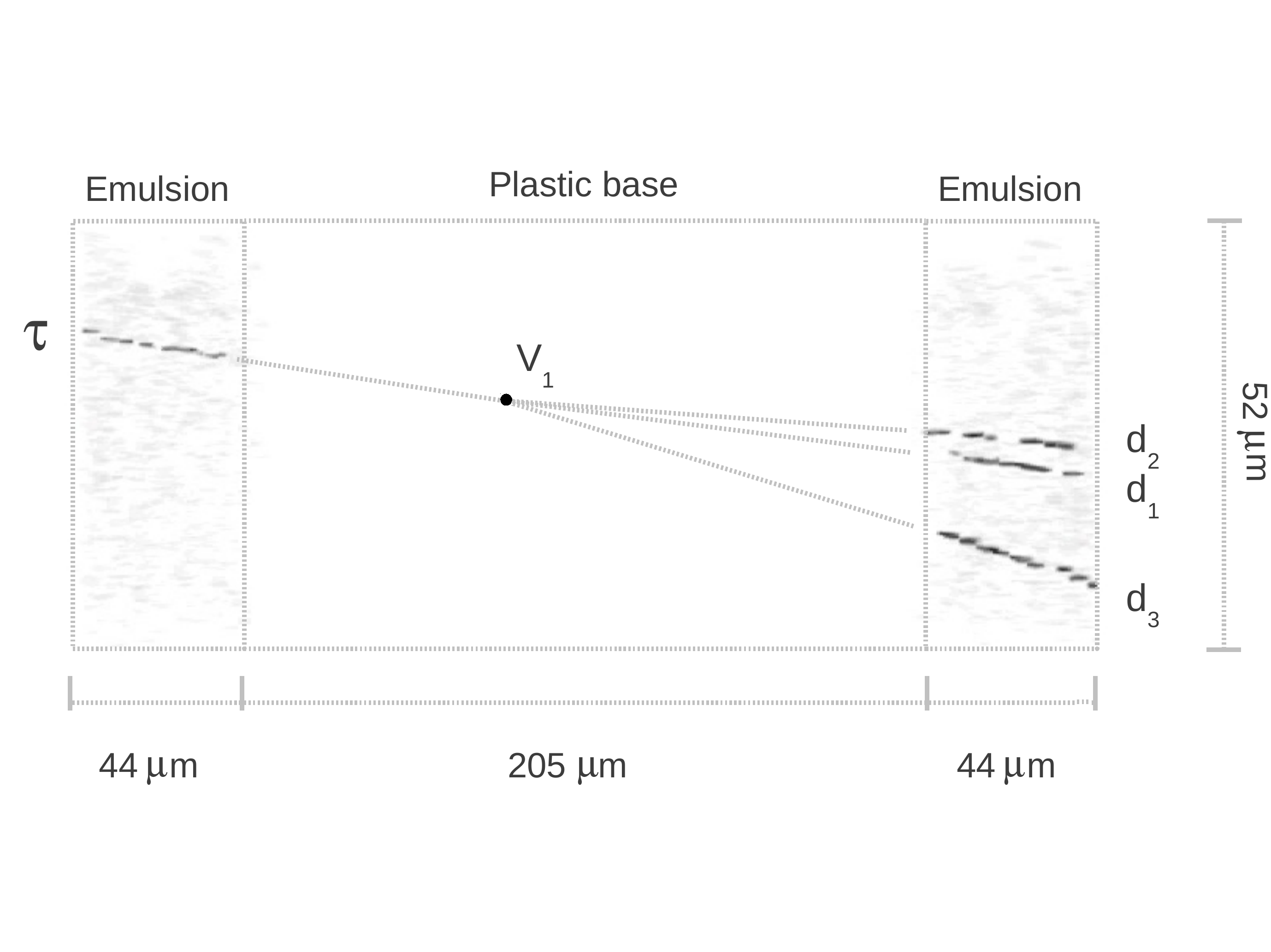}
\end{center}
\caption{Top-view display of the candidate $\tau\to 3h$ decay vertex
  ($V_1$ in Fig.~\ref{fig:fig_tau2a}).  The single grains observed by
  the optical microscope are visible in the emulsion layers (left-side
  and right-side rectangles). The directions of micro-tracks are
  extrapolated to the decay vertex in the plastic base region (central
  rectangle).
\label{fig:fig_tau2b}}
\end{figure}

The flight length of the $\tau$ lepton candidate is (1466 $\pm$
10)~$\mu$m and its decay occurs in the plastic base allowing excluding
with a high efficiency (above 99.8\% at 90\% C.L. up to $\tan
\theta=3$,~\cite{largeangle_a}) the emission of highly ionising
nuclear fragments.  The secondary vertex ($V_1$,
~Fig.~\ref{fig:fig_tau2a}) consists of three tracks (called $d_1$,
$d_2$, $d_3$).  A display of the reconstructed grains in the emulsion
layers is presented in Fig.~\ref{fig:fig_tau2b}. The impact parameters
of the decay products with respect to the reconstructed secondary
vertex are 1.3, 1.2 and 0.3~$\mu$m for $d_1$, $d_2$ and $d_3$
respectively.  After eye-inspection the background from instrumental
fake tracks or tracks due to Compton electrons is negligible.

\begin{figure}
\begin{center}
\includegraphics[width=8cm]{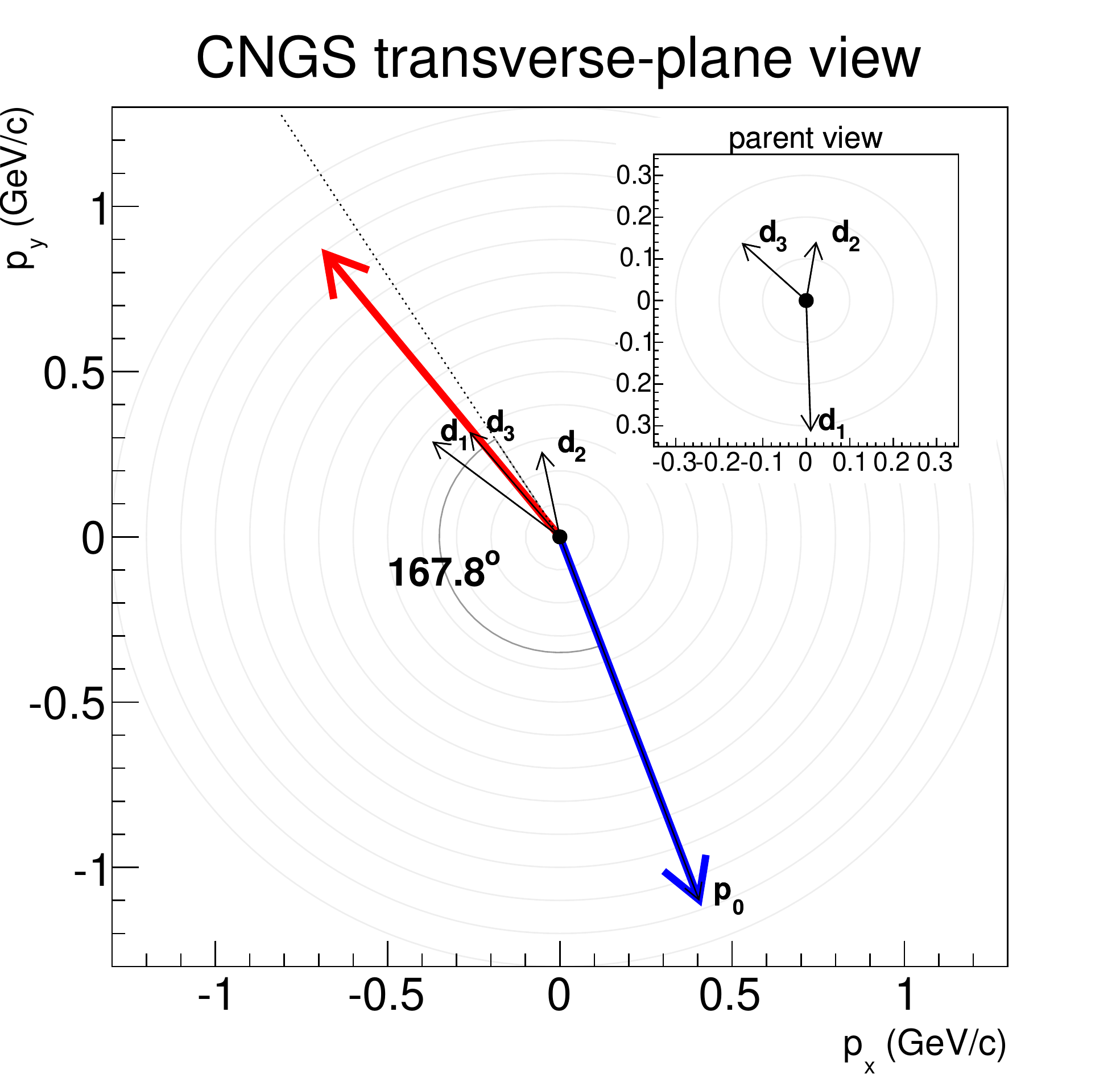}
\end{center}
\caption{Event display of $\nu_\tau$ candidate event B: \CNGS~transverse
  plane momentum balancing.  The red (blue) line represents the sum of
  the transverse momentum of the secondary (primary) vertex
  tracks. The dotted line marks the direction of the parent. The inset
  represents the transverse momenta of daughter tracks along the
  parent flight length direction.
\label{fig:fig_tau2_TP}}
\end{figure}

In the beam transverse plane the $\tau$ and the $p_0$ tracks form an
angle \mbox{$\Delta\phi_{\tau H} =(167.8 \pm1.1)^\circ$}
(Fig.~\ref{fig:fig_tau2_TP}).

In order to strongly constrain the hypothesis that the secondary
vertex could be a hadronic interaction, a search for nuclear fragments
has been performed both upstream and downstream of the vertex with an
automatic scanning up to $\tan \theta=3.5$ as well as by visual
inspection. No such fragment was found.

A search for \mbox{$\gamma$ conversions} has been performed up to
$\tan \theta=1$ for the 35 films (about 6~$X_0$) downstream of the
vertex yielding no candidates.

The analysis of each track to determine its nature and momentum is
described below:
\begin{itemize}
\item track $p_0$ has a measured momentum $p_{p_0} =
  (2.8^{+0.7}_{-0.7})$~GeV/$c$.  This track points towards the centre
  of the detector in the top-view such that its signature as a hadron
  is already well constrained using the target tracker information
  only.  Nevertheless the track has been followed in the downstream
  wall ($W_{1}$) where it is found to exit the brick to the
  side. Compatible tracks have neither been found in the adjacent
  brick in $W_{1}$ nor in downstream bricks (and CS doublets) up to
  $W_{3}$.  Gamma rays from a possible hadronic interaction were also
  searched for in two bricks in $W_{2}$, however, no $\gamma$ was
  found. This track is then assumed to interact in the dead material
  in-between two bricks. A muon with a momentum of the magnitude
  measured by MCS would be expected to travel from 26 to 44 brick
  layers before stopping.  This makes the muon hypothesis very
  unlikely: $D_{TFD}$ is 0.05 for this track.  Its projected slopes
  are $(0.155, -0.365)$, well inside the angular acceptance of the
  scanning. Even considering the possibility of having missed a muon
  track crossing the downstream emulsion detectors, the pattern in the
  scintillators does not allow the existence of such a track for more
  than about 7-8 brick walls.
\item track $d_1$ has slopes of \mbox{(-0.056, 0.101)} and a measured
  momentum $p_{d_1}=(6.6^{+2.0}_{-1.4})$~GeV/$c$. A hadronic
  interaction is detected in the emulsions of the brick in wall $W_4$
  (see Fig.~\ref{fig:fig_TTtau2}) producing two charged tracks
with slopes of \mbox{(0.234, 0.489)} and (0.034, -0.305). 
The signature of the interaction is also 
  indicated by the target tracker scintillators.
\item track $d_2$ has a slope of \mbox{(-0.041, 0.260)} and a measured
  momentum $p_{d_2}~=~(1.3^{+0.2}_{-0.2})$~GeV/$c$. It has not been
  found in $W_2$ or in the following walls corresponding to a
  range-momentum correlation parameter $D_{TFD}=0.25$.
\item track $d_3$ has a slope of \mbox{(-0.134, 0.220)} and a measured
  momentum $p_{d_3}~=~(2.0^{+0.9}_{-0.6})$~GeV/$c$. This track interacts
  in the brick containing the neutrino vertex, after 11 lead
  plates i.e. about 1.3~cm downstream ($V_2$ in Fig.~\ref{fig:fig_tau2a}). 
  The interaction occurs inside
  the emulsion 
resulting in a very clear signature.  The final state
  is composed of two charged tracks ($d_{3-1}$ and $d_{3-2}$) and four
  back-scattered nuclear fragments.
\end{itemize}
The scalar sum of the momenta of all the measured charged particles in
the event is 12.7$^{+2.3}_{-1.7}$~GeV/$c$.

\subsection{Summary of the two $\nu_\tau$ candidate events.}

\begin{figure}
\begin{center}
\includegraphics[width=7cm]{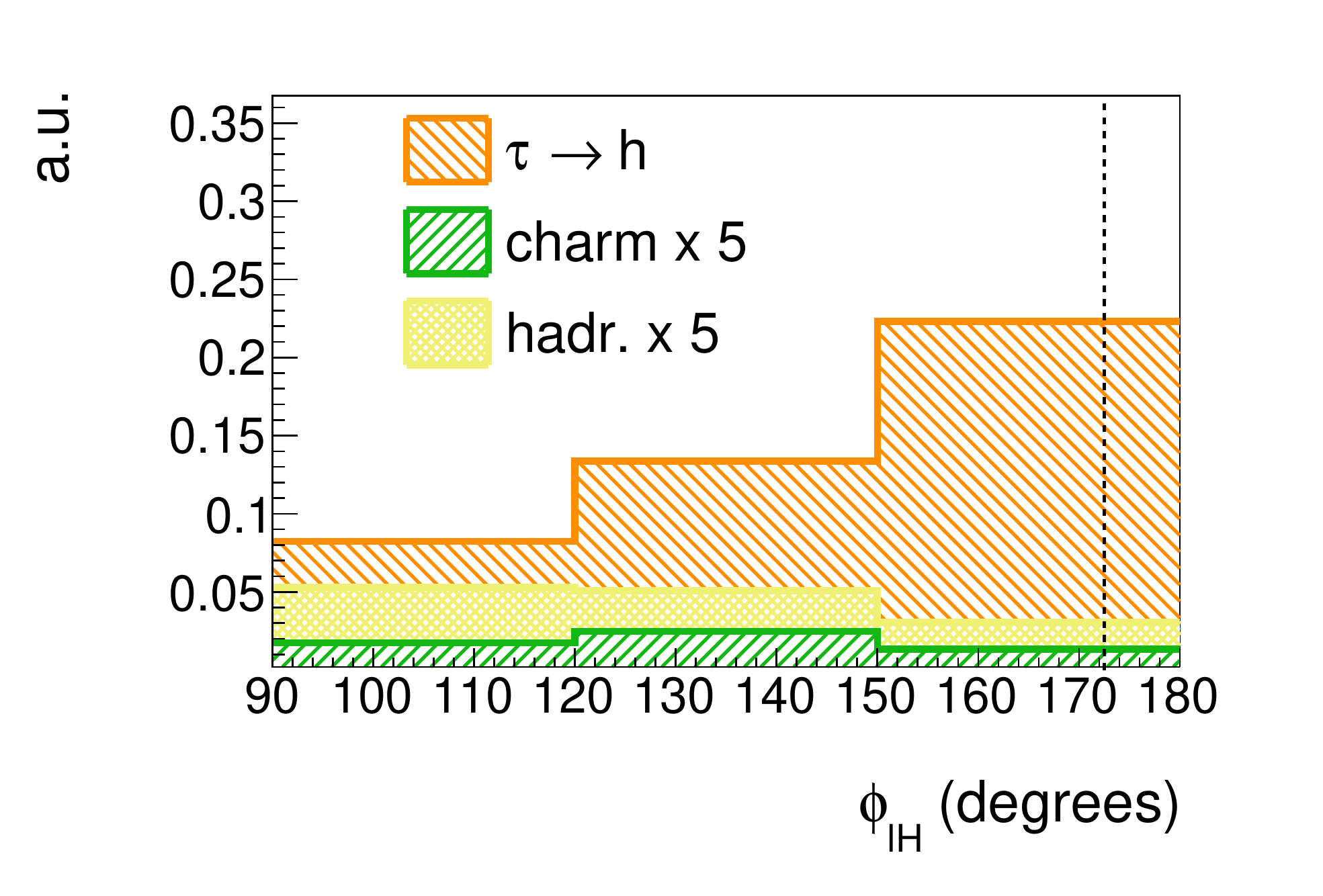}%
\includegraphics[width=7cm]{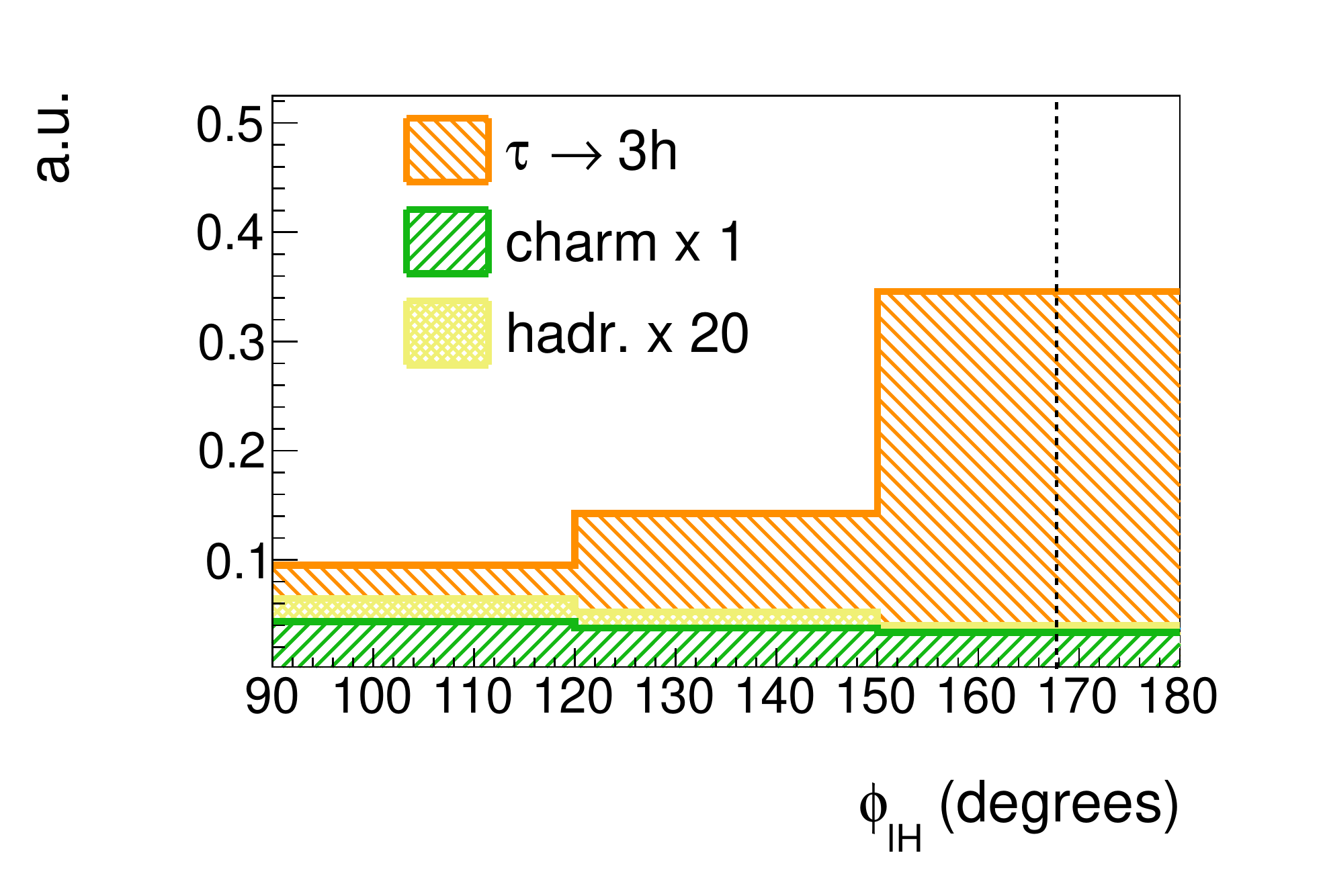}
\end{center}
\caption{Distributions of $\phi_{lH}$ for the $\tau\to 1h$ (left) and
  $\tau\to 3h$ (right) selections. The filled histograms represent in
  different colors (see the legend) the MC expectation for signal and
  background while the vertical lines represent the values measured
  for the $\tau$ candidates. The background and signal components are
  stacked and, for the sake of visualisation, backgrounds have been
  scaled up with the factors indicated in the
  legends.~\label{fig:finaldistr}}
\end{figure}

From the arguments developed in~\cite{ref1sttau,ref2ndoscpaper} and in
Sect.~\ref{sec:tau2desc}, it is concluded that the two events are
candidates of $\tau$ decays into the 1-charged hadron ($1h$) and
3-charged hadrons ($3h$) channels, respectively. Table~\ref{tab:tab1}
summarises the values taken by the kinematic variables for the two
candidates together with the cuts applied in the analysis.
\begin{table}
\centering
\small
\begin{tabular}{ccccccc}
\hline
Variable&$\tau\to 1h$ selection& candidate event A & $\tau \to 3h$ selection& candidate event B\\
\hline
$z_{dec}$ ($\mu$m) &$<$ 2600 & 435 $\pm$ 35 &idem& 1446 $\pm$ 10\\
$\phi_{lH}$ ($^\circ$)& $>90$ & $172.5 \pm 1.7$ &idem & $167.8 \pm 1.1$ \\
$\langle \theta_{kink} \rangle$ (mrad) & $>$ 20 & 41 $\pm$ 2& $<$ 500& 87.4 $\pm$ 1.5\\
$p_T^{miss}$ (GeV/$c$)& $<$ 1 & $0.57^{+0.32}_{-0.17}$& idem & $0.31 \pm 0.11$\\
$p_{2ry}$ (GeV/$c$)&$>2$&$12^{+6}_{-3}$&$>$ 3&  8.4 $\pm$ 1.7 \\
$p_{T,2ry}$ (GeV/$c$) & $>$ 0.3(0.6) &0.47$^{+0.24}_{-0.12}$ &/&/\\
$m_{min}$ (GeV/$c^2$)&/ & / &[0.5, 2]& 0.96 $\pm$ 0.13\\
$m$ (GeV/$c^2$)&/ & / &[0.5, 2]& 0.80 $\pm$ 0.12 \\
\hline
\end{tabular}
\caption{\label{tab:tab1} Selection criteria for $\nu_\tau$ candidate
  events and corresponding measured values.}
\end{table}
The distributions of $\phi_{lH}$, for the expected signal and
background components in the analysed sample are presented in
Fig.~\ref{fig:finaldistr} for the 1$h$ and 3$h$ channels
separately. The measured values in the data are indicated by vertical
lines.

\section{Significance of the observation}
\label{sec:multiv}

The significance of the observation of two candidates with the present
estimate of the backgrounds is addressed by considering the confidence
in the exclusion of the null hypothesis (i.e. having observed no
$\nu_\mu\to\nu_\tau$ oscillation signal). The probability that the two
events might be due to an upward fluctuation of the background is
defined as the integral of the Poisson distribution evaluated for
$n\geq 2$: $p=\sum_{n=2}^{\infty}{\mu^n\frac{e^{-\mu}}{n!}}$.  Using
for $\mu$ the sum of the background in the four channels (see
Fig.~\ref{fig:sigback}), the value \pValueSum is obtained,
corresponding to a \SigmaA exclusion (adopting the one-sided
definition).

In a similar manner the $p$-values of the two channels can be defined:
{\color{black}{\pValueOneh (\SigmaOneh)}} and
{\color{black}{\pValueThreeh (\SigmaThreeh)}}.  To combine them, an
approach based on generating a large number of pseudo-experiments has
then been followed.  For each of the four $\tau$ decay channels an
integer, $n_{i=1,...,4}$, is extracted from the Poisson distribution
of background.  The $p_i$~values are then calculated as above for each
extraction and the estimator of the results $p^\star$ is obtained by
taking their product $p^\star=p_1p_2p_3p_4$ (different choices of
estimator exist). The counting of the fraction of pseudo experiments
for which $p^\star\leq p_{1h}p_{3h}$ yields a significance of \SigmaB.
More elaborated estimates of the significance could be obtained using
a log-likelihood ratio analysis or other multi-variate
techniques. This approach will be addressed in future works.

\section{Conclusions and prospects}
The results of the $\nu_\tau$ appearance analysis on a pre-selected
sample of the neutrino interactions collected by the \OPERA~experiment
in the years 2008 to 2011 are reported.  A two-brick analysis was
performed on all 2008 and 2009 predictions while for 2010 and 2011 the
analysis was restricted to the most probable brick of all $0\mu$
events and of $1\mu$ events with $p_\mu<15$~GeV/$c$ of the 2010
sample.  A total of two $\nu_\tau$ candidates has been observed, one
in the $\tau\to 1h$ channel and one in the $\tau\to 3h$ channel. This
result is compatible with the expected amount of \ALLTHISfinal signal
events and \ALLBTHISfinal background events in all channels.  The good
agreement between data and MC, both for the location efficiencies of
$\nu_\mu$ events and for the detection of charmed particles, indicates
that the overall $\tau$ finding efficiencies are well understood.
Using the presently analysed sample the absence of a signal from
$\nu_\mu\to\nu_\tau$ oscillations (null hypothesis) is excluded at
\SigmaB.

In the near future, the analysis of the 2011 sample will be completed
using the same event selection as for 2010 as well as that of the 2012
sample currently in progress. Finally the search for events not found
in the HPB will be extended to second-priority bricks as it was done
for the 2008 and 2009 samples. The inclusion of 3$^{\rm{rd}}$ and
4$^{\rm{th}}$ bricks, which is also foreseen as a further step, will
bring an additional increase of the location efficiency.  The current
significance can be improved with the ongoing increase of the analysed
sample.

\section{Acknowledgments}

We thank \CERN~for the successful operation of the \CNGS~facility and
\textsc{INFN} for the continuous support given to the experiment
during the construction, installation and commissioning phases through
its \textsc{LNGS} laboratory. We warmly acknowledge funding from our
national agencies: Fonds de la Recherche Scientifique-\textsc{FNRS}
and Institut InterUniversitaire des Sciences Nucl\'eaires for Belgium,
\textsc{MoSES} for Croatia, \textsc{CNRS} and \textsc{IN2P3} for
France, \textsc{BMBF} for Germany, \textsc{INFN} for Italy,
\textsc{JSPS} (Japan Society for the Promotion of Science),
\textsc{MEXT} (Ministry of Education, Culture, Sports, Science and
Technology), \textsc{QFPU} (Global \textsc{COE} programme of Nagoya
University, Quest for Fundamental Principles in the Universe supported
by \textsc{JSPS} and \textsc{MEXT}) and Promotion and Mutual Aid
Corporation for Private Schools of Japan for Japan, \textsc{SNF}, the
University of Bern and \textsc{ETH} Zurich for Switzerland, the
Russian Foundation for Basic Research (grant no. 09-02-00300 a,
12-02-12142 ofim), the Programs of the Presidium of the Russian
Academy of Sciences Neutrino physics and Experimental and theoretical
researches of fundamental interactions connected with work on the
accelerator of \CERN, the Programs of Support of Leading Schools
(grant no. 3517.2010.2), and the Ministry of Education and Science of
the Russian Federation for Russia, the National Research Foundation of
Korea Grant No. 2011-0029457 for Korea and \textsc{TUBITAK}, the
Scientific and Technological Research Council of Turkey, for
Turkey. We are also indebted to \textsc{INFN} for providing
fellowships and grants to non-Italian researchers. We thank the
\textsc{IN2P3} Computing Centre (\textsc{CC-IN2P3}) for providing
computing resources for the analysis and hosting the central database
for the \OPERA~experiment. We are indebted to our technical
collaborators for the excellent quality of their work over many years
of design, prototyping and construction of the detector and of its
facilities.

%
%

\end{document}